\documentclass[reqno,11pt]{amsart}
\usepackage{amssymb, amsthm, amsfonts, amsgen, stmaryrd}
\usepackage{slashed}
\usepackage[english]{babel}
\usepackage{fullpage}
\usepackage[centertags]{amsmath}
\usepackage{indentfirst}
\usepackage{fancyhdr}
\usepackage[dvips]{graphicx}
\usepackage{psfrag}
\usepackage{enumerate}
\numberwithin{equation}{section}
\usepackage[latin1]{inputenc}

\usepackage{color}

\title[Some notes on Dirac operators  on spheres] {Some notes on Dirac operators on the   ${\rm S}^3$ and ${\rm S}^2$  spheres}
\date{18 september 2016}
\author{Fabio Di Cosmo}
\author{Alessandro Zampini}
\address{Dipartimento di Scienze Fisiche, Universit\`a di Napoli Federico II and INFN - Sezione di Napoli, Via Cintia - 80126 Napoli, Italy} 
\email{dicosmo@fisica.unina.it}
\address{ Mathematics Research Unit, University of Luxembourg, 6 Rue Richard Coudenhove - Kalergi - 1359  Cit\`e de Luxembourg - Luxembourg}
 \email{azampini@gmail.com}

\newcommand{\nn}{\nonumber}

\newcommand{\dd}{{\rm d}}

\newcommand{\cp}{\mathcal{P}}

\newcommand{\IC}{{\mathbb C}} 

\newcommand{\figureheight}{8cm}
\newcommand{\putfig}[2]{\begin{figure}[htp]
        \special{isoscale c:/itex/texfig/#1.wmf, \the\hsize \figureheight}
        \vspace{\figureheight}
        \caption{#2}\label{fig:#1}
        \end{figure}}
\newcommand{\pictureheight}{4cm}
\newcommand{\putpicture}[2]{\begin{figure}[htp]
        \special{isoscale c:/itex/texfig/#1.wmf, \the\hsize \pictureheight}
        \vspace{\pictureheight}
        \caption{#2}\label{fig:#1}
        \end{figure}}

\newcommand{\beqa}{\begin{eqnarray}}
\newcommand{\eeqa}{\end{eqnarray}}
\newcommand{\beq}{\begin{equation}}
\newcommand{\eeq}{\end{equation}}

\newcommand{\del}{\partial}

\newcommand{\R}{{\mathbb{R}}}
\newcommand{\Z}{{\mathbb{Z}}}

\newcommand{\C}{\mathbb{C}}

\newcommand{\TT}{\mathrm{T}}

\newcommand{\Spin}{\mathrm{Spin}}
\newcommand{\Pin}{\mathrm{Pin}}
\newcommand{\Cl}{\mathrm{Cl}}
\newcommand{\CL}{\C\mathrm{l}}
\newcommand{\D}{\mathcal{D}}
\newcommand{\bu}{\bar{u}}
\newcommand{\bv}{\bar{v}}

\newcommand{\ct}{\hat\theta}
\newcommand{\A}{\mathcal{A}}
\newcommand{\E}{\mathcal{E}}

\begin{document}

\thispagestyle{empty}

\begin{abstract}
We describe both the Hodge - de Rham and the spin manifold Dirac operator on the spheres ${\rm S}^3$ and ${\rm S}^2$,  following the formalism introduced by K\"ahler, and exhibit a complete spectral resolution for them in terms of suitably globally defined eigenspinors. 
\end{abstract}


\maketitle
\tableofcontents


\section{Introduction}
This paper outgrew from the lectures that one of the author was invited to deliver at the Workshop on \emph{Quantum Physics: foundations and applications} at the Centre for High Energy Physics in Bangalore in february 2016. The aim of these lectures was to outline, in a (hopefully)  pedagogical manner, the building blocks of the theory allowing to define a Dirac operator on a smooth riemannian manifold. 

The first question to be addressed was to describe that, given a riemannian smooth manifold $(M,g)$, it is possible to define several Dirac operators: they can all be consistently formulated as covariant  derivatives acting upon a space of spinors which can be defined as a suitable  module (over $C^{\infty}(M)$, the algebra of smooth functions on $M$) with respect to an action of the Clifford algebra corresponding to the metric tensor $g$. The section \ref{sec:KA}  of the paper introduces the notions of Clifford algebras and spin group at a global level, i.e.  corresponding to a metric tensor on a finite dimensional vector space. Upon recalling the notion of smooth manifold and bundle,  the section \ref{sec:bundles}  presents the notions of Clifford modules, bundles and (general) Dirac operators. A specific emphasis is given to the case of the Hodge - de Rham Dirac operator. It was  first written by K\"ahler and can be defined on any orientable riemannian manifold, and in such a case the relevant space of spinors can be realised as a subspace of the exterior algebra over $M$.  The section ends by recalling that  specific topological conditions allow to define and suitably characterise, within the set of all possible Dirac operators, the so called (Atiyah) spin manifold Dirac operator, which plays a prominent role in several branches of mathematics.  

The section \ref{S3} presents a detailed analysis of both the K\"ahler and the spin manifold Dirac operators on the spheres ${\rm S}^3$ and ${\rm S}^2$. The aim of this section is to show the differences between these two operators via some explicit examples: the emphasis is given to the fact that the Dirac operators can be studied, upon using some properties of the monopole bundle formulation,  within a global formalism that allows for a complete description of their spectra. 

The problem of defining a Dirac operator on a riemannian manifold has been so deeply studied from several different perspectives that the literature on it is ovewhelming. The first sections of this paper are strongly influenced by \cite{fof, VFG, graf,  landsman, lami,  varilly}, the last by \cite{balaimmi}.


\section{The K\"ahler-Atiyah algebra over a vector space}
\label{sec:KA}

Let $V$ be a finite $N$-dimensional vector space over $\R$.  Its tensor algebra  $(\TT(V), \otimes)$ is  the infinite dimensional vector space over $\R$ given by 
\beq
\label{tead}
\TT(V)\,=\,\oplus_{k\,=\,0}^{\infty}\,V^{\otimes k}
\eeq
where $V^{\otimes 0}\,=\,\R$ and $V^{\otimes k}\,=\,V\,\otimes\,\cdots\,\otimes\,V$ ($k$-times), together with the usual associative tensor product 
of its elements. Notice that this tensor algebra is $\Z$-graded, namely $V^{\otimes j}\,\otimes\,V^{\otimes k}\,\subset\,V^{\otimes j+k}$. 
Let $J\,\subset \,\TT(V)$ be the two-sided ideal generated by the elements $a\,\otimes a\,\in\,V^{\otimes 2}$ for any $a\,\in\,V$, that is $J\,=\,\TT(V)\,\otimes(a\,\otimes\,a)\otimes\,\TT(V)$. The exterior algebra $(\Lambda(V), \wedge)$ is the quotient 
\beq
\label{eade}
\Lambda(V)\,=\,\TT(V)/J
\eeq
where we denote by the wedge symbol, as usual, the associative product induced on $\Lambda(V)$ by   the tensor product in $\TT(V)$. Notice that, since $J$ is homogeneous with respect to the $\Z$-grading in $\TT(V)$, the quotient \eqref{eade} preserves the grading, and one  has the finite direct sum 
\beq
\Lambda(V)\,=\,\oplus_{k\,=\,0}^N \,\Lambda^k(V)
\eeq
where $\Lambda^{0}(V)\,=\,V,\,\,\Lambda^1(V)\,=\,V,\,\,\Lambda^k(V)\,=\,V^{\otimes k}/J$ for $k\,\geq\,2$, with $\Lambda^j(V)\wedge\Lambda^k(V)\,\subset\,\Lambda^{j+k}(V)$.   It is clear that    $\Lambda^{k}(V)\,=\,\{0\}$ for $k\,>\,N$ and $\dim\,\Lambda^k(V)\,=\,N!/(k!(N-k)!)$ so that, as a vector space,  $\dim\,\Lambda(V)\,=\,2^N$. The wedge product is indeed graded commutative, namely
\beq
a\wedge b\,=\,(-1)^{jk}b\wedge a
\label{gco}
\eeq
with $a\,\in\,\Lambda^j(V)$ and $b\,\in\,\Lambda^k(V)$. 

Let  a bilinear symmetric form $g$ be defined on $V$ 
and consider  the two-sided ideal $K\,\subset \,\TT(V)$ generated by elements $a\,\otimes\,a\,-\,g(a,a)$ for any $a\,\in\,V$, that is 
\beq
K\,=\,\TT(V)\,\otimes(a\,\otimes\,a\,-\,g(a,a))\otimes\,\TT(V).
\label{idcli}
\eeq
 The Clifford algebra 
$(\Cl(V,g), \vee)$ is the quotient 
\beq
\label{decl}
\Cl(V,g)\,=\,\TT(V)/K
\eeq
where the associative product $\vee$ is induced by the tensor product in $\TT(V)$. Since $K$ has inhomogeneous terms of  even degree in $\TT(V)$, the Clifford algebra naturally inherits a $\Z_2$-grading, so that we may write 
\beq
\label{gracl}
\Cl(V, g)\,=\,\Cl_{e}(V, g)\,\oplus\,\Cl_{o}(V, g)
\eeq
where $\Cl_e(V, g)$ ($\Cl_o(V,g)$) is given by the images of elements of even (odd) degree in $\TT(V)$. The Clifford product satisfies the rule
\beq 
\label{cld}
a\vee b\,+\,b\vee a\,=\,2g(a,b)
\eeq
for any $a,b\,\in\,V$. This relation allows to easily prove that:
\begin{enumerate}
\item as a vector space, $\dim\,\Cl(V, g)\,=\,2^N$. The map 
\begin{align}
\Lambda(V)\quad&\rightarrow\quad\Cl(V,g) \nonumber \\
v_1\wedge\cdots\wedge v_k\quad&\mapsto\quad\frac{1}{k!}\,\sum_{\sigma\,\in\,S_k}(-1)^{\pi(\sigma)}v_{\sigma(1)}\vee\cdots\vee v_{\sigma(k)}
\label{qua}
\end{align}
(with $S_k$ the permutation group of $k$ elements, and $\pi(\sigma)$ the parity of the permutation $\sigma$) is a vector space isomorphism between the exterior algebra over $V$ and a Clifford algebra over $V$ (irrespective of the bilinear symmetric form $g$ on it).
\item 
if the bilinear form $g$ is degenerate,  the Clifford product coincides with the exterior product on images of elements in $\TT(V)$ generated by vectors $v\,\subset\,V_0\,\subseteq\,V$ where $V_0$ is the null subspace of $g$. In what follows we shall restrict our attention to Clifford algebras corresponding to non degenerate symmetric bilinear forms $g$, i.e. (not necessarily positive definite) metric tensors.

\end{enumerate}

The metric tensor $g$ allows to define suitable contraction map on $\Lambda(V)$. For each element $v\,\in\,V$ one defines the linear map  $i_v\,:\,\Lambda^k(V)\,\to\,\Lambda^{k-1}(V)$ by setting
\begin{align}
&i_v(\lambda)\,=\,0\qquad\forall\,\lambda\,\in\,\Lambda^0(V)\sim\R, \nonumber \\
&i_v(w)\,=\,g(v,w), \qquad \forall\,w\,\in\,V\label{co1}
\end{align}
and extending the action of $i_v$ upon all elements in $\Lambda(V)$ by requiring it to be linear and to  satisfy a graded Leibniz rule with respect to the wedge product. 
Given the basis $\{e_j\}_{j\,=\,1, \ldots, N}$ for $V$ and its dual basis $\{\epsilon^k\}_{k\,=\,1, \ldots, N}$ for $V^*$, with $\epsilon^k(e_j)\,=\,\delta^k_j$,  one writes $g_{ab}\,=\,g(e_a, e_b)$ so that one has $g\,=\,g_{ab}\epsilon^a\otimes\epsilon^b$ or equivalently $g\,=\,g^{ab}e_a\otimes e_b$ on $(V^*)^*$ with $g^{ab}g_{bc}\,=\,\delta^a_c$. If one recalls the vector space isomorphism between the exterior algebra $\Lambda(V)$ and the Clifford algebra $\Cl(V,g)$, one can write the Clifford product on $\Lambda(V)$ by
\beq
\label{clv}
 \phi\vee\omega\,=\,\sum_s\,\frac{(-1)^{\tiny{\left(\begin{array}{c}s \\ 2 \end{array}\right)}}}{s!}g^{{a_1}{b_1}}\,\cdots\,g^{{a_s}{b_s}}(\check{\gamma}^s\{i_{e_{a_1}}\,\cdots \,i_{e_{a_s}}\,\phi\})\wedge\{i_{e_{b_1}}\,\cdots\,i_{e_{b_s}}\,\omega\}, 
\eeq
where $\phi,\,\omega\,\in\,\Lambda(V)$ and $\check{\gamma}(\omega)\,=\,(-1)^k\omega$ if $\omega\,\in\,\Lambda^k(V)$.  This means that the exterior algebra $\Lambda(V)$ is an associative unital algebra with respect to both the wedge and the Clifford product. The set $(\Lambda(V), g,\wedge, \vee)$ is usually called  the K\"ahler-Atiyah algebra over $(V, g)$.\footnote{A direct  proof of \eqref{clv} is in \cite{ka}, while the following  argument, which can be used toward proving \eqref{clv},
 is described in \cite{fof}. Consider the map $\Phi\,:\,V\,\to\,\mathrm{End}(\Lambda(V))$ defined on a basis of $V$ via (with $\omega\,\in\,\Lambda(V)$)
\beq
\label{deF}
\Phi(e_a)\,:\,\omega\qquad\mapsto\qquad e_a\wedge\omega\,+\,i_{e_a}\omega.
\eeq
Such a map satisfies the identity
\beq
\label{clmap}
\Phi(v)\Phi(w)\,+\,\Phi(w)\Phi(v)\,=\,2g(v,w)
\eeq
for any pair $v,w\,\in\,V$. By the universality of the Clifford algebra, this map extends to a unique algebra homomorphism $\Phi\,:\,\Cl(V,g)\,\to\,\mathrm{End}(\Lambda(V))$, and upon evaluating $\Phi$ at $1\,\in\,\Lambda(V)$ one gets an isomorphism between $\Cl(V,g)$ and $\Lambda(V)$ whose explicit form is \eqref{clv}.}

It is well known that a metric tensor $g$ on $V$ provides the exterior algebra
$\Lambda(V)$ a natural scalar product.  Elements in $\Lambda^k(V)$ and $\Lambda^{k^{\prime}}(V)$ are defined orthogonal if $k\,\neq\,k^{\prime}$, while one sets (recall \eqref{co1}):
\beq
\label{scaproe}
\langle e_{a_1}\wedge\ldots\wedge e_{a_s}\,\mid\,e_{b_1}\wedge\ldots\wedge e_{b_s}\rangle\,=\,i_{e_{a_s}}\,\ldots\,i_{e_{a_1}}\,(e_{b_1}\wedge\ldots\wedge e_{b_s})
\eeq
Such a scalar product on $\Lambda(V)$ allows to define a scalar product on $\Cl(V,g)$. One sets
\beq
\label{scaprocl}
\langle e_{a_1}\vee\ldots\vee e_{a_s}\,\mid\,e_{b_1}\vee\ldots\vee e_{b_s}\rangle_{\Cl}\,=\,
\langle 1 \mid e_{a_s}\vee\ldots\vee e_{a_1}\vee e_{b_1}\vee\ldots\vee e_{b_s}\rangle
\eeq
where the rhs is defined via \eqref{scaproe}.

Given the Clifford algebra $\Cl(V, g)$, one defines the corresponding ${\rm Pin}(V,g)$ as a subgroup of the group of units of $\Cl(V,g)$, namely  
\beq
\label{pin}
{\rm Pin}(V,g)\,=\,\{v_{a_1}\vee\cdots\vee v_{a_j} \,:\,g(v_{a_k}, v_{a_k})\,=\,\pm1\}.
\eeq
For each $v\,\in\,\Pin(V,g)\cap V$ one may define the map (an adjoint action) ${\rm Ad}_v\,:\,V\,\to\,{\rm End}(V)$ given by
\beq
x\,\in\,V, \qquad{\rm Ad}_v(x)\,=\,-v\vee x\vee v^{-1}\,=\,x\,-\,2\,\frac{g(x,v)}{g(v,v)}\,v:
\label{rexv}
\eeq
this map reflects $x$ with respect to a hyperplane perpendicular to $v$. The map ${\rm Ad}$ can be clearly multiplicatively  extended to 
the whole $\Pin(V,g)$ group, giving an adjoint action of $\Pin(V,g)$ upon  $V$. It is well known that the following sequence of group homomorphisms 
\beq
\label{esP}
1\quad\longrightarrow\quad\{\pm1\}\quad\longrightarrow\quad\Pin(V,g)\quad\stackrel{{\rm Ad}}{\longrightarrow}\quad{\rm O}(V,g)\quad\longrightarrow\quad 1
\eeq
is exact: the $\Pin(V,g)$ group is a double covering of the orthogonal group ${\rm O}(V,g)\,\subset\,{\rm End}(V)$ given by linear maps $T$ satisfying the relation $g(Tv, Tw)\,=\,g(v,w)$ for any $v,w,\,\in\,V$. The $\Spin(V,g)$ group is defined as
\beq
\label{despi}
\Spin(V,g)\,=\,\Pin(V,g)\,\cap\,\Cl_{e}(V,g):
\eeq
it is given by the even part of the $\Pin(V,g)$ group.
One has the following exact sequence of group homomorphisms:
\beq
\label{essP}
1\quad\longrightarrow\quad\{\pm1\}\quad\longrightarrow\quad\Spin(V,g)\quad\stackrel{{\rm Ad}}{\longrightarrow}\quad{\rm SO}(V,g)\quad\longrightarrow\quad 1.
\eeq
The $\Spin(V,g)$ is a double covering of the special orthogonal ${\rm SO}(V,g)$ group, that is the subgroup of orthogonal transformations $T\,\in\,{\rm O}(V,g)$ with ${\rm det}\,T\,=\,1$. 

We close this section by considering the Clifford algebra defined on a finite dimensional complex vector space. Since a complex quadratic form has no signature, every finite dimensional complex quadratic vector space is isomorphic to an orthogonal direct sum $(\C^N, g)\,=\,\C^r(0)\oplus\C^s(1)$  where $\C^r(0)$ is the $r$ dimensional kernel of $g$, while $\C^s(1)$ is the $s$-dimensional complex vector subspace of $V$ with an orthonormal  basis $\{v_{a_j}\}_{j\,=\,1,\ldots,s}$ giving   $g(v_{a_k}, v_{a_p})\,=\,\delta_{kp}$. If $(V,g)$ is a $N$-dimensional real vector space, it can be complexified: one has $V_{\C}\,=\,V\otimes_{\R}\C$ and $g_{\C}(zv, z^{\prime}v^{\prime})\,=\,zz^{\prime}g(v,v^{\prime})$ with $v,v^{\prime}\in\,V$ and $z,z^{\prime}\,\in\,\C$. It is easy to see that
\beq
\label{complc}
\Cl(V_{\C}, g_{\C})\,\simeq\,\Cl(V,g)\,\otimes_{\R}\C.
\eeq
In the following we shall denote $\CL(V,g)\,=\,\Cl(V,g)$. With respect to such a complexified Clifford algebra, one defines
the analogue of the ${\rm Spin}(V,g)$ group as 
\beq
\label{spinc}
{\rm Spin}^{c}(V, g)\,=\,\{v_{a_1}\,\vee\,\cdots\,\vee\,v_{a_{2k}}\,:\,v_{a_j}\,\in\,V_{\C}, \,g_{\C}(v_{a_j}^*,v_{a_j})\,=\,1\}
\eeq
with $v^*$ denoting the complex conjugate of $v\,\in\,V_{\C}$. The ${\rm Ad}$ map defined in \eqref{rexv} can be immediately extended to the complex case, so one has the short exact sequence
\beq
\label{esspc}
1\quad\longrightarrow\quad{\rm U}(1)\quad\longrightarrow\quad\Spin^c(V,g)\quad\stackrel{{\rm Ad}}{\longrightarrow}\quad{\rm SO}(V,g)\quad\longrightarrow\quad 1.
\eeq

\section{Clifford algebra bundles and Dirac operators}
\label{sec:bundles}

\subsection{An inner calculus over an exterior algebra}
\label{ssec:inner}
Let $M$ be a real $N$-dimensional smooth  manifold, and $\pi\,:\,E\,\to\,M$ a rank $k$ real vector bundle on it, with typical fiber a $k$-dimensional real vector space $E_x$. 
We denote by $\E\,=\,\Gamma(M, E_{\C})$ the set of complex valued sections of the bundle, which is a finitely generated projective symmetric\footnote{Symmetry for the $\A$-bimodule $\E$ means that the left and the right multiplication of $\A$ upon $\E$ coincide.}    $\A\,=\,C^{\infty}(M)$-bimodule. Some refinements of the celebrated theorem by Serre and Swan prove that the category of smooth vector bundles over a smooth manifold $M$ is equivalent to the category of finitely generated projective modules over $\A$. 
A (smooth) metric structure on the vector bundle $\pi\,:\,E\,\to\,M$  is given by a real non degenerate symmetric $\A$-bilinear form $\gamma\,:\,\E\times \E\,\to\,\A$.    By real form it is intended that  $\gamma(\sigma,\sigma^{\prime})$ is a real function on $M$ if $\sigma, \sigma^{\prime}$ are real sections in $\E$; by symmetry and bilinearity it is intended that $\gamma(h \sigma, \sigma^{\prime})\,=\,\gamma(\sigma^{\prime}, h\sigma)\,=\,h\gamma(\sigma, \sigma^{\prime})$ with $h\,\in\,\A$; the non degeneracy clearly amounts to say that the condition $\gamma(\sigma, \sigma)\,=\,0$ for any real  $\sigma\,\in\,\E$   implies $\sigma\,=\,0$. 
An hermitian structure on $E$ is defined by 
\beq
\label{hefib}
\langle\sigma\mid\sigma^{\prime}\rangle\,=\,\gamma(\sigma^*,\sigma^{\prime}).
\eeq
 It is immediate to see that, if $E\,=\,TM$ -- the tangent bundle to $M$ -- then such a metric  structure can be identified with a metric tensor $g$ on $M$. 
The metric structure on the vector bundle $E$ allows to consider each typical fiber $E_x$ equipped with a metric $g_x$ (here $x\,\in\,M$), and then the corresponding Clifford algebra $\CL(E_x, g_x)$ which is a $2^k$-dimensional vector space.  The set $\CL(E, g)\,=\,\amalg_{x\,\in\,M}\CL(E_x, g_x)$ can be given the structure of a smooth vector bundle with typical fiber $\CL(E_x, g_x)$, and it is called the Clifford bundle corresponding to $\pi\,:\,E\,\to\,(M, g)$. 
If $E\,=\,T^*M$ -- the cotangent bundle on $M$ --  then from \eqref{qua} one sees that 
\begin{align}
&\Gamma(M, \Cl(T^*M, g))\,\simeq\,\Lambda(M)\,=\,\oplus_{j\,=\,1}^N\Lambda^j(M), \nn \\
\label{extka}
&\Gamma(M, \CL(T^*M, g))\,\simeq\,\Lambda(M)\otimes_{\R}\C\,=\,\oplus_{j\,=\,1}^N\Lambda^j(M)\otimes_{\R}\C,
\end{align}
i.e. the set of sections of the Clifford bundle is isomorphic, as a $\A$-bimodule, to the set of exterior forms defined on $M$. This isomorphism allows to extend the Clifford product from  $\Gamma(M, \CL(T^*M, g))$ to $\Lambda_{\C}(M)$. 

Assume  $M$ is given a suitable smooth atlas and $\{x^j\}_{j=1,\ldots,N}$ is the corresponding local coordinate system, with $\del_j\,=\,\del/\del x^j$ the local basis for the set of vector fields $\mathfrak{X}(M)\,=\,\Gamma(M, T_{\C}M)$ and $\{\dd x^j\}_{j=1,\ldots,N}$ its dual local basis for exterior 1-forms, so that one can write $g\,=\,g^{ab}\del_a\otimes\del_b$ for the metric tensor. The Clifford product on $\Lambda(M)$ takes the form
\beq
\label{clm}
 \phi\vee\phi^{\prime}\,=\,\sum_s\,\frac{(-1)^{\tiny{\left(\begin{array}{c}s \\ 2 \end{array}\right)}}}{s!}g^{{a_1}{b_1}}\,\cdots\,g^{{a_s}{b_s}}(\check{\gamma}^s\{i_{a_1}\,\cdots \,i_{a_s}\,\phi\})\wedge\{i_{b_1}\,\cdots\,i_{b_s}\,\phi^{\prime}\}, 
\eeq
where $\phi,\,\phi^{\prime}$ are elements in $\Lambda(M)$, one has  $\check{\gamma}(\phi)\,=\,(-1)^k\phi$ for $\phi\,\in\,\Lambda^k(M)$ ($\check{\gamma}$ is the degree operator) and $i_{a}\,=\,i_{\del_a}$ is the contraction operator on $\Lambda(M)$\footnote{Given a vector field $X\,\in\,\mathfrak{X}(M)$, the contraction map $i_X\,:\,\Lambda^k(M)\,\to\,\Lambda^{k-1}(M)$ is the natural  generalization of the contraction map introduced in the previous pages on a vector space. Such a map is linear and uniquely defined by:
\begin{align}
i_X(f)\,=\,0,& \qquad\qquad\forall \,f\,\in\,\Lambda^0(M) \nn\\
i_X(\phi)\,=\,\phi(X),& \qquad\qquad\forall\,\phi\,\in\,\Lambda^1(M) \nn \\
i_X(\phi\wedge\phi^{\prime})\,=\,(i_X(\phi))\wedge\phi^{\prime}\,+\,(-1)^k\phi\wedge i_X(\phi^{\prime}),& \qquad\qquad\forall\,\phi\,\in\,\Lambda^k(M).
\label{coM}
\end{align}
Notice that $i_Xi_Y\,=\,-i_Yi_X$ for any pair of vector fields $X,Y\,\in\,\mathfrak{X}(M)$. 
On the exterior algebra the exterior derivative operator $\dd\,:\,\Lambda^k(M)\,\to\,\Lambda^{k+1}(M)$ is uniquely defined by
\begin{align}
\dd f(X)\,=\,X(f)\,=\,L_Xf, &\qquad\qquad\qquad\forall\,f\,\in\,\Lambda^0(M), \,X\,\in\,\mathfrak{X}(M) \nn \\
\dd(\phi\wedge\phi^{\prime})\,=\,(\dd\phi)\wedge\phi^{\prime}\,+\,(-1)^k\phi\wedge \dd\phi^{\prime},& \qquad\qquad\qquad\forall\,\phi\,\in\,\Lambda^k(M) \nn \\
\dd^2\phi\,=\,0,&\qquad\qquad\qquad\forall\,\phi\,\in\,\Lambda(M),
\label{dde}
\end{align} 
where $L_Xf$ denotes the Lie derivative of the function $f$ along the vector field $X$. One denotes by $(\Lambda(M), \wedge, \dd, i_X)$ the exterior algebra over $M$ equipped with its natural exterior differential calculus: the Cartan identity -- that is $L_X\,=\,\dd\,i_X\,+\,i_X\dd$ when acting upon elements in $\Lambda(M)$ -- holds.}.
One clearly has
\begin{align}
\dd x^a\vee\dd x^b&=\,\dd x^a\wedge\dd x^b\,+\,g^{ab}, \nn \\
\dd x^a\vee\dd x^b\,+\,\dd x^b\vee\dd x^a&=\,2g^{ab}.
\label{esv}
\end{align}
What we have described above shows that  $\mathfrak{K}(M,g)\,=\,\amalg_{x\,\in\,M}\CL(T_x^*M, g_x)$ 
can be given the structure of a smooth bundle over $M$ of Clifford algebras, whose sections can be identified with $\Gamma_{\mathfrak{K}(M, g)}\,=\,\Lambda(M)\otimes_{\R}\C\,=\,\Lambda_{\C}(M)$. The expression \eqref{clm} shows that we may then consider such a set of sections 
\beq
\label{kamg}
\Gamma_{\mathfrak{K}(M, g)}\,=\,(\Lambda_{\C}(M), \vee, \wedge, \dd, i_X)
\eeq
 as the 
 K\"ahler-Atiyah algebras over $(M, g)$. The unital algebraic structure $(\Lambda_{\C}(M), \vee)$ is called, after K\"ahler, an \emph{inner} product over an exterior algebra.

\subsection{Clifford modules and Dirac operators}
\label{ssec:clidi}
Given the  smooth real finite dimensional manifold equipped with a metric tensor $(M,g)$, we say that  a finitely generated projective bimodule $(\E, \,\langle~~\mid~~\rangle)$ over $\A\,=\,\C^{\infty}(M)$, with $\langle~~\mid~~\rangle$ an hermitian structure on it, 
 is a Clifford module if the two following conditions are satisfied:
 \begin{enumerate}
 \item 
  $\E$ is a  left module with respect to a left action of the Clifford (\emph{inner}) algebra structure $(\Lambda_{\C}(M), \vee)$, that is a linear $\A$-homomorphism $\nu\,:\,(\Lambda_{\C}(M), \vee)\,\to\,{\rm End}(\E)$ exists;
   \item  the elements in the range of the homomorphism  $\nu$ provide  selfadjoint maps on $(\E, \langle~~\mid~~\rangle)$, that is 
 \beq 
 \label{sacl}
 \langle\sigma\mid\nu(\theta)\sigma^{\prime}\rangle\,=\,\langle\nu(\theta^*)\sigma\mid\sigma^{\prime}\rangle
\eeq
with $\theta\,\in\,\Lambda_{\C}(M), \,\sigma, \sigma^{\prime}\,\in\,\E$.
\end{enumerate}
In order to introduce the notion of Dirac operator on a Clifford module, we recall the notion of covariant derivative (connection) on a finitely generated projective module. Let $\E$ be a finitely generated projective  $\A$-module, 
that is a $\A$-linear  map $p\,:\,\A^{\otimes s}\,\to\,\A^{\otimes s}$ (with $s$ a suitable integer) exists, such that  $\E\,=\,p\,\A^{\otimes s}$ with $p\,=\,p^{\dagger}\,=\,p^2$. A  connection on $\E$  is a map
\beq
\delta\,:\,\Lambda^k(M)\,\otimes_{\A}\E\,\to\,\Lambda^{k+1}(M)\otimes_{\A}\E
\label{dena}
\eeq
such that a graded Leibniz rule is satisfied,
\beq
\label{legra}
\delta(\theta\,\sigma)\,=\,(\dd \theta)\,\sigma\,+\,(-1)^k\theta\,\delta\sigma
\eeq
for any $\sigma\,\in\,\E$ and $\theta\,\in\,\Lambda^k(M)$. A connection is completely characterized by its restriction $\delta\,:\,\E\,\to\,\Lambda^1(M)\otimes_{\A}\E$, satisfying 
$$
\delta(f\sigma)\,=\,(\dd f)\,\sigma\,+\,f\,\delta\sigma
$$
(with $f\,\in\,\A$) and then extended by using the Leibniz rule. The composition 
$$
\delta^2\,=\,\delta\,\circ\,\delta\,:\,\Lambda^k(M)\otimes_{\A}\E\,\to\,\Lambda^{k+2}(M)\otimes_{\A}\E
$$ 
is $\Lambda(M)$-linear, the restriction $F\,=\,\delta^2$ to $\E$ is the curvature corresponding to the connection $\delta$, with $\delta^2(\theta\,\sigma)\,=\,\theta\,\delta^2(\sigma)$,   $\theta\,\in\,\Lambda(M)$ and $\sigma\,\in\,\E$.  Connections always exist on a projective module; the set of connections  is an affine space, since 
 any two connections $\delta$ and $\delta^{\prime}$ differ by a matrix valued 1-form element, i.e.
$\delta-\delta^{\prime}\,\in\,\mathbb{M}_s(\A)\otimes_{\A}\Lambda^1(M)$, so one can write $\delta\,=\,\delta_0\,+\,A$ with $\delta_{0}\,=\,p\,\dd$ (it is the Grassmann connection) and $A\,\in\, \Lambda^1_{\C}(M)\,\otimes_{\A}\,\mathbb{M}_s(\A)$ with $p\,A\,=\,A\,p$. The element $A$ is referred to as the vector potential of the connection $\delta$.
The operator $\delta_{X}\,=\,i_X\,\otimes\,1\,:\,\E\,\to\,\E$ evaluates the covariant derivative $\delta$ along the vector field $X\,\in\,\mathfrak{X}(M)$.  

Let us assume that  $\{X_a\}_{a\,=\,1, \ldots, N}$ gives a (local) basis for $\mathfrak{X}(M)$ and $\{\theta^a\}_{a\,=\,1,\ldots,N}$ its dual (local) basis on $\Lambda^1(M)$. Along such a basis, the metric tensor on $M$ can be written as $g\,=\,g^{ab}X_a\otimes X_b$.   
Let $\{\epsilon_j\}_{j\,=\,1, \ldots, s}$ be a basis for the free $\A$-(bi)module $\A^{\otimes s}$, with $\E\,=\,\{f^j\epsilon_j\,:\,f^j\,\in\,\A, \,p(f^j\epsilon_j)\,=\,f^j\epsilon_j\}$. A covariant derivative on $\E$ is completely characterised   
by
\beq
\label{cepo}
\delta(\epsilon_j)\,=\,\theta^lA_{lj}^{\,\,\,q}\epsilon_q
\eeq
with $A_{lj}^{\,\,\,q}\,\in\,\A$ and $j,q\,\in\,1,\ldots, s$ while $l\,\in\,1, \ldots, N\,=\,{\rm dim}\,M$, so that  
\beq
\delta(f^j\epsilon_j)\,=\,(\dd f^j)\otimes_{\A}\epsilon_j\,+\,f^j\delta(\epsilon_j)\,=\,\theta^l\{X_lf^q\,+\,A_{lj}^{\,\,\,q}f^j\}\epsilon_q.
\label{povet}
\eeq
If the projective module $\E$ is given an hermitian structure, then  a connection is called riemannian when
\beq
\delta_X\langle\sigma\mid\sigma^{\prime}\rangle\,=\,\langle\delta_X\sigma\mid\sigma^{\prime}\rangle\,+\,\langle\sigma\mid\delta_X\sigma^{\prime}\rangle.
\label{riecon}
\eeq
If $(\E, \langle~\mid~\rangle)$ is a Clifford module over $(M,g)$, a connection $\delta$ on it is called compatible with the Clifford structure if it is riemannian and if
\beq
\label{cpati}
\delta_X(\theta\vee\sigma)\,=\,(\nabla_X\theta)\vee\sigma\,+\,\theta\vee(\delta_X\sigma)
\eeq 
(with $X\,\in\,\mathfrak{X}(M), \, \theta\,\in\,\Lambda_{\C}(M)$ and $\sigma\,\in\,\E$) where we have denoted via the Clifford product symbol the left (Clifford) action of $\Lambda_{\C}(M)\,\simeq\,\Gamma_{\mathfrak{K}(M.g)}$ upon $\E$ 
$$
\theta\vee\sigma\,=\,\nu(\theta)\sigma
$$
and $\nabla$ is the Levi-Civita connection\footnote{It is well known -- as a specific case of the general theory described above --  that the Levi-Civita connection corresponding to $(M,g)$ is defined on $\mathfrak{X}(M)\,=\,\Gamma(M, TM)$ and $\Lambda^1(M)\,=\,\Gamma(M, T^*M)$ via 
\begin{align}
&\nabla\,:\,\Lambda^1(M)\,\to\,\Lambda^1(M)\,\otimes_{\A}\,\Lambda^1(M), \nn \\
&\nabla\,:\,\mathcal{X}(M)\,\to\,\Lambda^1(M)\,\otimes_{\A}\,\mathcal{X}(M),
\end{align} 
with (compare such expression with \eqref{povet}) 
\begin{align}
\nabla(\dd x^a)&=\,-\dd x^s\,\otimes\,\Gamma^a_{sb}\dd x^b, \nn \\
\nabla(\del_a)&=\,\dd x^s\,\otimes\,\Gamma^b_{sa}\del_b
\label{nabd}
\end{align}
where $\Gamma^a_{bc}$ are the Christoffel symbols of the 
connection.  
\beq
\label{Cs}
\Gamma_{ji}^m\,=\,\Gamma_{ij}^m\,=\,\frac{1}{2}g^{mk}\left(\del_jg_{ki}\,+\,\del_ig_{kj}\,-\,\del_kg_{ij}\right).
\eeq
The action of $\nabla$ is extended to $\Lambda(M)$ via requiring it to satisfy the Leibniz rule with respect to the wedge product.  } corresponding to $g$ acting on $\Lambda_{\C}(M)$. Given an hermitian finitely generated projective module $\E$ over $(M,g)$, each compatible connection defines a Dirac operator upon $\E$, given via
\beq
D\,\sigma=\,\theta^a\,\vee\,\delta_{X_a}\sigma\,=\,\nu(\theta^a)\,\delta_{X_a}\sigma
\label{dirde}
\eeq
It is clear that $\nu(\theta^a)\,\in\,\A\,\otimes_{\A}\mathbb{M}_s(\C)$. The matrices $\nu(\theta^a)\,=\,\gamma^{(a)}$ are the gamma-matrices corresponding to the Clifford module $\E$ (notice that they depend also on the choice of the frame $\{\theta^a\}_{a=1,\ldots, N}$), and they satisfy the relation
\beq
\label{mga}
\gamma^{(a)}\gamma^{(b)}+\gamma^{(b)}\gamma^{(a)}\,=\,2g^{ab}.
\eeq
From \eqref{povet} one can write the action of the Dirac operator \eqref{dirde} as
\begin{align}
\label{dioper}
D(f^j\epsilon_j)&=\,\{X_lf^q\,+\,A_{lj}^{\,\,\,q}f^j\}\nu(\theta^l)\epsilon_q
\\
&=\,\{X_lf^q\,+\,A_{lj}^{\,\,\,q}f^j\}(\gamma^{(l)})^t_{q}\epsilon_t: 
\end{align}
upon the components $f^j\,\in\,\A$ of $\sigma\,\in\,\E$ one has
\beq
\label{acdco}
D(f^j)\,=\,(\gamma^{(l)})^j_q\{X_lf^q\,+\,A_{ls}^{\,\,\,q}f^s\}. 
\eeq
For a given smooth manifold equipped with a metric tensor $(M, g)$ it is then possible to define several Dirac operators. In the following pages we shall focus on two specific examples. 

\subsection{The Hodge-de Rham Dirac operator}
\label{ssec:HD}

Following \cite{ka}, we review how the Hodge - de Rham operator can be defined on the exterior algebra over a manifold $(M,g)$, and how it can be formulated as a suitable Dirac operator.  For any  $\phi\,\in\,\Lambda^k(M)$, its exterior covariant derivative is defined via
\beq
\label{eD}
\mathfrak{D}\phi\,=\,\dd x^a\,\wedge\,\nabla_{a}\phi,
\eeq
where $\nabla_a\,=\,\nabla_{\del_a}$. The action of the K\"ahler  operator is defined as:
\beq
\label{did}
\mathcal{D}\phi\,=\,\dd x^a\,\vee\,\nabla_a\phi.
\eeq 
The action of this operator can immediately be extended to $\Lambda_{\C}(M)$. From the identification \eqref{kamg} it is evident that $\Lambda_{\C}(M)$ is a Clifford module with respect to the Clifford (inner)  product on itself;  the  K\"ahler operator 
$\mathcal{D}$ is the Dirac operator \eqref{dirde} corresponding to the Levi-Civita connection on $\Lambda_{\C}(M)$ with $\delta_{X_a}\,=\,\nabla_a$.
From \eqref{clm} one immediately sees that
\beq
\label{dve}
\D\phi\,=\,\mathfrak{D}\phi\,+\,g^{ab}i_b\nabla_a\phi.
\eeq
A direct computation shows that $\mathfrak{D}\phi\,=\,\dd\phi$, so one has
\beq
\label{die}
\D\phi\,=\,\dd\phi\,+\,g^{ab}i_b\nabla_a\phi:
\eeq
if $\phi\,\in\,\Lambda^k(M)$, the image $\D\phi$ has a component of degree $k+1$ and a component of degree $k-1$. Notice that $\D\phi$ has a well defined parity with respect to the $\Z_2$ grading of $\Lambda(M)$. One can prove that, for $\phi\,\in\,\Lambda^k(M)$, 
\beq
\label{difi}
\D\phi\,=\,\dd\phi\,+\,(-1)^{N(k-1)}\star\dd\star\phi
\eeq
where one has introduced the Hodge duality operator corresponding to the metric $g$, i.e. a $\A$-bimodule map $\star\,:\,\Lambda^k(M)\,\to\,\Lambda^{N-k}(M)$ defined on a basis by
\beq
\label{hod}
\star(\dd x^{a_1}\wedge\cdots\wedge\dd x^{a_k})\,=\,g^{a_1b_1}\,\cdots\,g^{a_kb_k}i_{b_k}\,\cdots\,i_{b_{1}}\,\mu,
\eeq
with  $\mu\,=\,|g|^{1/2}\dd x^1\wedge\,\cdots\,\wedge\,\dd x^N$ is the invariant volume form 
corresponding to the metric $g$ (here $|g|$ denotes the determinant of the matrix $g_{ab}$).  The  relation 
\beq
\label{hos}
\star^2(\phi)\,=\,(-1)^{k(N-k)}({\rm sgn}\,g)\phi
\eeq
holds, for $\phi\,\in\,\Lambda^k(M)$, with $({\rm sgn}\,g)$ the signature of the metric tensor $g$.
Via a metric $g$ on a manifold, one can introduce a scalar product among exterior forms. For homogeneous $\phi$ and $\phi^{\prime}$ with the same degree we set 
\beq
\label{scp}
\langle\phi\mid\phi^{\prime}\rangle\,=\,\int_{M}\phi\,\wedge\,\star\phi^{\prime},
\eeq
where the integration measure on $M$ comes from the volume form $\mu$ previously defined. 
Homogeneous  forms $\phi, \phi^{\prime}$ are defined orthogonal if their degrees are different (this amounts to say that $\Lambda^k(M)$ is orthogonal to $\Lambda^{k^{\prime}}(M)$ for $k\,\neq\,k^{\prime}$). Notice that such a scalar product is the natural generalization to the exterior algebra $\Lambda(M)$ of the scalar product introduced in \eqref{scaproe}\footnote{\label{notap} It is easy to prove that, if the exterior algebra $\Lambda(M)$ is a free ${\mathcal A}$-module (which amounts to eventually consider only a local description for a manifold which is not globally parallelasible) then the expression \eqref{scp} defines a Hodge duality once a non degenerate hermitian scalar product is given on the whole $\Lambda(M)$.}. 
Given any $\alpha\,\in\,\Lambda^k(M)$ and $\beta\,\in\,\Lambda^{k+1}(M)$, upon defining 
$\dd^*\,=\,(-1)^{N(k-1)+1}\star\dd\star\,:\,\Lambda^{k}(M)\,\to\,\Lambda^{k-1}(M)$ one proves that
\beq
\label{agd}
\langle\dd\alpha\mid\beta\rangle\,=\,\langle\alpha\mid\dd^*\beta\rangle\,+\,\int_M\dd(\alpha\wedge\star\beta).
\eeq
This relation shows that the operator $\dd^*$ can then be considered -- under suitable conditions on its domain --  the adjoint of $\dd$ in $\Lambda(M)$ with respect to the scalar product \eqref{scp}. The action \eqref{difi} of the Dirac operator
can be cast in the form
\beq
\label{diag}
\D\phi\,=\,\dd\phi\,-\,\dd^*\phi:
\eeq
from \eqref{hos} we have that $\dd^{*}\dd^*\,=\,0$, making it   immediate to prove the following relation
\beq
\label{laD}
\D^2\phi\,=\,-(\dd\dd^*+\dd^*\dd)\phi\,=\,(-1)^{kN}\{\star\dd\star\dd\phi\,+\,(-1)^N\dd\star\dd\star\phi\}
\eeq
for $\phi\,\in\,\Lambda^k(M)$. The square of the Dirac operator gives a version of the Hodge - de Rham Laplacian on $\Lambda(M)$\footnote{The following relation is usually referred to as a Green's identity,
\beq
\label{gri}
\langle\D^2\phi\mid\phi^{\prime}\rangle\,-\,\langle\phi\mid\D^2\phi^{\prime}\rangle\,=\,\int_M\dd\{(\dd^*\phi^{\prime})\wedge\star\phi\,+\,\phi^{\prime}\wedge\star\dd\phi\,-\,\phi\wedge\star(\dd\phi^{\prime})\,-\,(\dd^*\phi)\wedge\star\phi^{\prime}
\}
\eeq
with $\phi, \,\phi^{\prime}\,\in\,\Lambda^k(M)$, and allows to analyze the condition under which the operator $\D^2$ is self-adjoint.
An identity generalizing to the Dirac operator $\D$ the expression \eqref{agd} 
valid for $\dd$ exists. In order to describe  it, one considers exterior forms 
\beq
\alpha\,=\,\sum_{k=0}^N\alpha_{(k)}, \qquad
\beta\,=\,\sum_{k=0}^N\beta_{(k)}
 \label{inh}
\eeq 
with $\alpha_{(k)},\beta_{(k)}\,\in\,\Lambda^k(M)$ and defines
\beq
\Lambda^N(M)\,\ni\,(\alpha, \beta)\,=\,\sum_{k=0}^N\alpha_{(k)}\wedge\star\beta_{(k)}.
\label{ksc}
\eeq
From  \eqref{scp} 
one sees that the quantity $\int_{M}(\alpha,\beta)$ is the scalar product between $\alpha$ and $\beta$ defined in \eqref{inh} as the sum of inhomogeneous terms.  One then defines
\begin{align}
\Lambda(M)\,\otimes\,\Lambda(M)\qquad&\to\qquad\Lambda^{N-1}(M) \nn \\
\alpha\,\otimes\,\beta\qquad&\mapsto\qquad(\alpha, \beta)_1\,=\,i_k(\dd x^k\vee\alpha,\beta)
\label{sc1}
\end{align}
which gives
\beq 
(\alpha, \beta)_1\,=\,\sum_{k=0}^N\{\alpha_{(k)}\wedge\star\beta_{(k+1)}\,+\,\beta_{(k)}\wedge\star\alpha_{(k+1)}\},
\label{s1}
\eeq
showing that $(\alpha, \beta)_1\,=\,(\beta, \alpha)_1$. It  is indeed easy to prove that
\beq
\label{Dsa}
(\alpha, \D\beta)\,+\,(\D\alpha, \beta)\,=\,\dd(\alpha,\beta)_1,
\eeq
thus generalizing the identity \eqref{agd}.}.

\subsection{ Algebraic Spinors}
\label{algspinor}

It is easy to see that the action of the Dirac - K\"ahler operator upon $\Lambda_{\C}(M)$  defined in  \eqref{did} is highly reducible. The aim of the present section is to review a formalism developed in \cite{graf}, which allows to introduce the notion of irreducible spinors with respect to such action. 
Since the Clifford algebra $(\Lambda(M), \vee)$ is semi simple (simple for even $N$)\footnote{An interesting description of the representation theory for both real and complex  Clifford algebras is in \cite{fof}.} , its finite dimensional irreducible representations are given by its minimal (left) ideals $I\,\subset\,\Lambda(M)$, with $\Lambda(M)\,\vee\,I\,\subset\,I$.
The decomposition of this algebra into minimal ideals can be characterized by a spectral set $P_j$ of $\vee$-idempotents in $(\Lambda(M), \vee)$, i.e. a set of elements satisfying:
\begin{enumerate}
\item $\sum_jP_j\,=\,1$
\item $P_j\vee P_k\,=\,\delta_{jk}P_j $
\item the rank of $P_j$ is minimal (non trivial), where the rank of $P_j$ is given by the dimension of the range of the  $\Lambda(M)$-morphism $\psi\,\mapsto\,\psi\vee P_j$.
\end{enumerate}
Under these conditions one has that 
\beq 
\label{dI}
I_j\,=\,\{\psi\,\in\,\Lambda(M)\,:\,\psi\vee P_j=\psi\}
\eeq
 and $\Lambda(M)\,=\,\oplus_jI_j$.
Elements in $I_j$ are called {\it spinors} with respect to the idempotent $P_j$. Different sets of spinors, that is different ideals $I_j, I_k$, are called equivalent if an element $\epsilon\,\in\,\Pin(M, g)$ exists such that 
\beq 
\label{eqP}
P_j\,=\,\epsilon\vee P_k\vee\epsilon^{-1}.
\eeq
Different sets of spinors are called strongly equivalent if the condition \eqref{eqP} is satisfied by an element $\epsilon\,\in\,\Spin(M,g)$. Notice that this notion of equivalence among spinors amounts to say that two sets $I_j, I_k$
of spinors are equivalent if and only if for each element $\psi_j\,\in\,I_j$ there is an element $\epsilon\,\in\,\Pin(M,g)$ such that $\psi_j\vee\epsilon\,\in\,I_k$.

The action of the Dirac operator $\D$ 	defined in \eqref{difi} is meaningful on a set of spinors, i.e. $\D$ maps elements of $I_j$ into elements in $I_j$ with $I_{j}$ the left ideals of the Clifford algebra $(\Lambda(M), \vee)$, if and only if the condition
\beq
\label{compD}
P_j\vee\nabla_aP_j\,=\,0
\eeq
for any $\nabla_{a}\,=\,\nabla_{\del_a}$ holds. 

Let $\{U_a\,\subseteq\,M\}$ be the local charts of a smooth atlas for $M$, with local coordinates $\{x^a\}_{a\,=\,1,\ldots,N}$. 
Let $P$ be a suitable projector for the Clifford algebra $(\Lambda(U), \vee)$ whose range is denoted by $I_P\,\subset\,\Lambda(U_a)$. We denote by $\{w_i\}_{i=1,\ldots, s}$ with $s\,=\,{\rm rank}\,I_P$ a basis for it: $I_P$ is a free module over $C^{\infty}(U_a)$ given by
\beq 
I_P\,\sim\,\C^s\,\otimes\,C^{\infty}(U_a)\,\ni\,\psi\,=\,\psi^iw_i
\label{demoIP}
\eeq 
with $\psi^i\,\in\,C^{\infty}(U_a)$.
We have
\beq
\label{ddxx}
\dd x^a\,\vee\,w_i\,=\,(\gamma^{(a)})^j_iw_j.
\eeq
This equation defines the $s$-dimensional gamma matrices $\gamma^{(a)}$ which generate the  action of the Clifford algebra $\Cl(U_a, \vee)$ upon $I_P$.  The relation \eqref{dioper} can be written as:
\begin{align}
\D\psi\,=\,\D(\psi^iw_i)&=\,(\nabla_k\psi^i)\,\dd x^k\vee\,w_i\,+\,\psi^i\,\dd x^k\vee(\nabla_k w_i) \nn \\
&=(\del_k\psi^j\,+\,\psi^iA_{ki}^{\,\,j})\,\dd x^k\vee w_j \label{diess}\\
&=(\del_k\psi^j\,+\,\psi^iA_{ki}^{\,\,j})\,(\gamma^{(k)})^{s}_{j}w_s
\label{diesp}
\end{align}
where we have defined 
\beq
\label{dedie}
\nabla_kw_i\,=\,A_{ki}^{\,\,j}w_j:
\eeq
notice that such a definition is meaningful since the relation \eqref{compD} holds.
Such an identification shows that the Hodge - de Rham Dirac operator -- as defined by K\"ahler in \eqref{did} --  acts on the space of spinors $I_P$ as a covariant derivative, with a vector potential given by \eqref{dedie}

\subsection{The spin manifold Dirac operator}
\label{ssec:spi}

Given the equivalence between finitely generated projective modules over $\A\,=\,\C^{\infty}(M)$ and finite rank smooth vector bundles over $M$, the problem of describing Clifford bundles as  spaces of sections of  bundles associated  to  suitable  principal  bundles over $M$ is studied  in detail in several textbooks, see for example  \cite{fried, lami}. 
We assume  $(M,g)$ to be an orientable $N$-dimensional smooth manifold equipped with the metric tensor $g$ and   introduce the orthonormal frame bundle $P_{{\rm SO}(g)}(M)$, i.e. a principal ${\rm SO}(g)$-bundle over $M$. 
A spin structure for the  manifold $(M,g)$ is a principal ${\rm Spin}(g)$-bundle $P_{{\rm Spin}(g)}(M)$ with basis $M$ such that a 2-sheeted covering 
\beq 
\label{2sc}
\pi\,:\,P_{{\rm Spin}(g)}(M)\,\to\,P_{{\rm SO}(g)}(M)
\eeq
satisfying $\pi(p\,\gamma)\,=\,\pi(p)\,\pi_0(\gamma)$ for $p\,\in\,P_{{\rm Spin}(g)}, \,\gamma\,\in\,{\rm Spin}(g)$ and $\pi_0\,:\,{\rm Spin}(g)\,\to\,{\rm SO}(g)$ is the 2-covering from \eqref{essP}. We recall that a spinor structure on $(M,g)$ exists if and only if $w_2(M)\,=\,0$, i.e. the second Stiefel-Whitney class on $M$ (rigorously  on $TM$) is trivial.
By $\Spin(g)$ we mean the spin group corresponding to the metric vector space $(V, g)\,=\,(T^*_xM, g_x)$ given by the typical fiber of $T^*M$.   Let $W$ be a finite dimensional left (rank $s$) $\Cl(V,g)$-module, i.e. let us assume that  an algebra homomorphism  $\rho\,:\,\Cl(V,g)\,\to\,{\rm Aut}(W)$ exists. By a real spinor bundle we mean 
\beq
\label{respi}
S(M)\,=\,P_{{\rm Spin}(g)}\,\times_{\rho}\,W,
\eeq
that is the vector bundle associated to the ${\rm Spin}(g)$-principal bundle $P_{{\rm Spin}(g)}(M)$ via the representation $\rho$ of $\Cl(V,g)$ on $W$. Notice that the map $\rho$ defines a representation of ${\rm Spin}(g)$ on $W$ since ${\rm Spin}(g)\,\subset\,\Cl_0(V,g)$. If we consider the complex Clifford algebra $\Cl_{\C}(V,g)$ and a complex vector space $W$ we get the complex spinor bundle
\beq
\label{cespi}
S_{\C}(M)\,=\,P_{{\rm Spin}(g)}\,\times_{\rho}\,W.
\eeq
We denote by $\Gamma(S)$ the set of sections of the vector bundle $S(M)$. The action $\rho$ of $\Cl(V,g)$ upon $W$ induces an action of $\Cl(V,g)$ upon $\Gamma(S)$, the Clifford product being obviously represented as the  matrix multiplication in ${\mathbb M}(W)$. 

In the example of the K\"ahler - Dirac operator defined above one is considering $\CL(M,g)$ as a left $\A$-module over itself  whose action is given by the left multiplication. One easily sees that this  is the associated bundle
\beq 
\label{asscli}
S_{\C}(M)\,=\,P_{\Spin(g)}\times_{\lambda}\CL(V,g)
\eeq
with $\lambda$ denoting the left multiplication. The elements in the  $\A$-bimodule $\Gamma(S)$ (or $\Gamma(S_{\C})$)
are called spinors. The associated bundle is called irreducible if the algebra homomorphism $\rho$ defined above is irreducible.

We close this section by describing how to define the well known spin Dirac operator on a manifold $(M,g)$ which admits a spin structure. On the space of section $\Gamma(S)$ corresponding to a real spinor bundle (like in \eqref{respi}), 
with  $\psi\,\in\,\Gamma(S)$, the operator 
\beq
\delta^{\nabla}\,:\,\psi\quad\mapsto\quad(\dd\,+\,\frac{1}{4}\,\gamma^{(a)}\gamma^{(b)}\,\omega_{ab})\,\psi\,\in\,\Lambda^1(M)\otimes\Gamma(S)
\label{nabco}
\eeq 
 is the spin connection covariant derivative on $\Gamma(S)$, with (we adopt the same notations  as in the subsection \ref{algspinor} for a local chart system, and further assume that within such a coordinate system the metric is cast into normal form, so that $g\,=\,\eta_{ab}\dd x^a\otimes\dd x^b$ with $\eta_{ab}\,=\,\pm1$ depending on the signature of $g$)
 \beq
 \label{spicoome}
 \omega_{ab}\,=\,\dd x^r\,\eta_{as}\,\Gamma_{rb}^{\,\,\,s}
 \eeq 
the spin connection 1-form, given by the lift to $P_{{\rm SO}(g)}(M)$ of the Levi Civita connection on $TM$, so that $\omega_{ab}\,=\,-\omega_{ba}$.  We write (locally) $\Gamma(S)\,\ni\,\psi\,=\,\psi^iw_i$ with $\{w_i\}$ a basis for $W$ and $\psi^i\,\in\,C^{\infty}(U_a)$, we have
\beq 
\label{dnabd}
\delta^{\nabla}\psi\,=\,\dd x^r\,\delta^{\nabla}_r\,\psi.
\eeq 
This expression defines  the spin connection covariant derivative $\delta^{\nabla}_r\,:\,\Gamma(S)\,\to\,\Gamma(S)$ along the coordinate directions $\del_r$:
\beq
\label{detirr}
\delta^{\nabla}_r\,\psi\,=\,(\del_r\psi^i\,+\,\frac{1}{4}\,\psi^i\eta_{as}\Gamma_{rb}^{\,\,\,s}\gamma^{(a)}\gamma^{(b)})w_i
\eeq
The corresponding Dirac operator is  $\slashed D\,:\,\Gamma(S)\,\to\,\Gamma(S)$: 
\beq
\slashed D\,\psi\,=\,\gamma^a\,\delta^{\nabla}_a\psi:
\label{ASDirac}
\eeq
this is  the spin connection Dirac operator, following the definition given by Atiyah and Singer. The analysis above shows that the action \eqref{ASDirac} comes,  via the representation $\rho$, from the action of the  operator 
\beq
\label{spigeomD}
\slashed D\,\psi\,=\,\theta^a\,\vee\,\delta^{\nabla}_a\,\psi
\eeq
upon a set of spinors $\Gamma(S)$ which is a suitable left $\Cl(V, g)$-module. 
The action of the  spin Dirac operator defined in \eqref{spigeomD} closely resembles that of the Hodge - de Rham Dirac operator defined \eqref{diess}, when one identifies $I_P$ with $\Gamma(S)$ and the corresponding actions of the Clifford algebra generated by the differential 1-forms.  
It is immediate to see that the spin manifold Dirac operator acts upon $I_P$ as a covariant derivative characterized by a vector potential  given by
\beq
\label{cospige}
\Lambda^1(U_a)\,\otimes\,{\mathbb M}_s(\C)\,\ni\,\delta^{\nabla}w_i\,=\,\theta^a\,A_{ai}^{\,\,\,j}w_j\,=\,
\frac{1}{4}\,\theta^a\eta_{ks}\Gamma_{ab}^{\,\,\,s}(\gamma^{(k)}\gamma^{(b)}w_i)
\eeq
on an orthonormal basis for $\Lambda^1(U_a)$. It is clear that the spin connection covariant derivative defined in  \eqref{detirr}  and the covariant derivative defined in  \eqref{diesp} may differ, when the metric tensor has a non vanishing curvature\footnote{This point is elucidated in \cite{nicolae}.}.
One can see that \cite{tucker}, upon the identification $I_P\,\sim\,\Gamma(S)$, the relation
\beq
\label{diffconne}
(\delta^{\nabla}_a\,-\,\nabla_a)w_i\,=\,\frac{1}{4}\,w_i\,\vee\,\{(\eta_{sb}\nabla_a\theta^s)\wedge\theta^b\}
\eeq
holds. It is moreover possible to prove that the two covariant derivative above are not gauge related. 

The spin manifold Dirac operator on a compact riemannian smooth manifold $(M,g)$ can be indeed interestingly characterised by suitable conditions \cite{varilly, VFG}. Define 
\beq
\label{defibb}
B\,=\,\left\{\begin{array}{lcl} \Gamma(M, \Cl(T^*M)), & ~ & {\rm if}\,\,\dim\,M=2m \\ 
\Gamma(M, \Cl_{o}(T^*M)), & ~ & {\rm if}\,\,\dim\,M=2m+1.
\end{array}
\right.
\eeq
By comparing such a definition with \eqref{extka} one sees that $B$  is a vector bundle whose fibers are simple algebras (with respect to the Clifford $\vee$ product) of finite dimension $2^{2m}$. One can prove that
\begin{enumerate}[(1)]
\item the pair $(M,g)$ has a ${\rm Spin}^c(g)$-structure if and only if there exists a finite projective hermitian $\A$-module $S$ carrying a self-adjoint action (i.e. this action satisfies \eqref{sacl}) of $B$ such that ${\rm End}_{\A}(S)\simeq B$; 
\item a ${\rm Spin}^c(g)$-structure on $(M,g)$ can be refined to a ${\rm Spin}(g)$-structure if and only if an antilinear endomorphism $C$ (called a charge conjugation map) of $S$ exists, such that\footnote{Notice that the map $\chi$ is defined by $\chi(v_1\vee\ldots\vee v_j)=(-1)^j\bar{v_1}\vee\ldots\vee\bar{v_j}$.}
\begin{enumerate}
\item $C(\sigma f)=C(\sigma)f$ for any $\sigma\in S$ and $f\in\A$, 
\item $C(u\sigma)=\chi(\bar{u})C(\sigma)$ for any $\sigma\in S$ and $u\in B$,
\item $C$ is antiunitary, i.e. $\langle C\sigma\mid C\sigma^{\prime}\rangle=\langle\sigma^{\prime}\mid\sigma\rangle$ for any $\sigma, \sigma^{\prime}\in S$, 
\item $C^2=\pm1$ on $S$ for $M$ connected.
\end{enumerate}
\end{enumerate}
The module $S$ is a Clifford module with respect to $B$. If the pair $(S, C)$ defines a ${\rm Spin}(g)$-structure for $(M,g)$, then there is only one connection $\delta\,:\,S\,\to\,\Lambda^1(M)\otimes_{\A}S$ which is riemannian (see \eqref{riecon}), compatible (see \eqref{cpati}) with the Clifford algebra structure given by $B$ and real, i.e.   
$[\delta_X, C]=0$ for any real vector field $X$ on $M$. Such a connection turns out to be the spin manifold Dirac connection that can be defined, if and only if $w_2(M)=0$, in terms of principal ${\rm Spin}(g)$-bundles over $(M,g)$. The equivalence of the two approaches is described in \cite{plymen, schroeder}. Notice that such a uniqueness for the spin manifold connection strongly depends on the (Morita) condition ${\rm End}_{\A}(S)\simeq B$, which amounts, in a principal bundle language, to restrict our attention only on spinors given as sections of associated bundle corresponding to irreducible representations of the relevant Clifford algebras.

\section{Dirac operators on a class of spheres}
\label{S3}

In this section we focus our attention to the spheres ${\rm S}^3$ and ${\rm S}^2$. 
We start by  realising the sphere ${\rm S}^3$ as the ${\rm SU}(2)$ Lie group manifold, i.e. the set of elements 
\beq
\label{sug}
\gamma\,=\,\left(\begin{array}{cc} u & -\bar{v} \\ v & \bar{u}\end{array}\right), 
\qquad\bar{u}u+\bar{v}v=1.
\eeq
Our aim is to describe the global differential calculus on it.
From the  relation 
$$
[\sigma_a, \sigma_b]=2i\epsilon_{ab}^{\,\,\,\,c}\sigma_c
$$ 
 expressing the commutator of the Pauli matrices, we set $T_a=-i\sigma_a/2$ as a basis of the Lie algebra $\mathfrak{su}(2)$, with
\beq
\label{cT}
[T_a, T_b]\,=\,\epsilon_{ab}^{\,\,\,\,c}T_c.
\eeq
The Maurer-Cartan 1-form
\beq
\label{dth}
\gamma^{-1}\dd \gamma\,=\,T_a\otimes\theta^a
\eeq
implicitly defines a left-invariant  basis $\{\theta^{a}\}_{a=1,\ldots,3}$ of $\Lambda^1({\rm SU}(2))$, with 
\begin{align}
&\theta^x\,=\,i(u\dd v-v\dd u+\bar{v}\dd\bar{u} - \bar{u}\dd \bar{v}), \nn \\
&\theta^y\,=\,u\dd v-v\dd u-\bv\dd\bu+\bu\dd\bv, \nn \\
&\theta^z\,=\,2i(\bu\dd u+\bv\dd v)\,=\,-2i(u\dd\bu+v\dd\bv), \label{tfl}
\end{align}
while the following  vector fields
\begin{align}
&L_x\,=\,\frac{i}{2}\,(\bv\del_u-v\del_{\bu}-\bu\del_v+u\del_{\bv}), \nn \\
&L_{y}\,=\,\frac{1}{2}\,(-\bv\del_u-v\del_{\bu}+\bu\del_v+u\del_{\bv}), \nn \\
&L_z\,=\,-\frac{i}{2}\,(u\del_u-\bu\del_{\bu}+v\del_v-\bv\del_{\bv}) \label{livf}
\end{align}
give the dual left-invariant basis for $\mathfrak{X}({\rm SU}(2))$, naturally closing the Lie algebra structure given by $[L_a,L_b]=\epsilon_{ab}^{\,\,\,\,c}L_c$. Notice that one usually defines the ladder operators
\begin{align}
&L_-\,=\,\frac{1}{\sqrt{2}}\,(L_x-iL_y)\,=\,\frac{i}{\sqrt{2}}\,(\bv\del_u-\bu\del_v), \nn \\
&L_+\,=\,\frac{1}{\sqrt{2}}\,(L_x+iL_y)\,=\,\frac{i}{\sqrt{2}}(u\del_{\bv}-v\del_{\bu})
\label{lpm}
\end{align}
with 
\beq
\label{lapm}
[L_-,L_+]=iL_z, \qquad[L_-,L_z]=-iL_-, \qquad[L_+,L_z]=iL_+,
\eeq
and dual 1-forms given by
\begin{align}
&\theta^-\,=\,\frac{1}{\sqrt{2}}\,(\theta^x+i\theta^y)\,=\,i\sqrt{2}(u\dd v-v\dd u), \nn \\
&\theta^+\,=\,\frac{1}{\sqrt{2}}\,(\theta^x-i\theta^y)\,=\,i\sqrt{2}(\bv\dd\bu-\bu\dd\bv).
\label{l1pm}
\end{align}
The differential calculus is finally characterized by 
\beq
\dd\theta^a\,=\,-\frac{1}{2}f^{\,\,\,a}_{bc}\theta^b\wedge\theta^c
\label{mceq}
\eeq
in terms of the Lie algebra structure constants along a given left invariant basis,
with
\beq
\label{Lth}
L_a\theta^b\,=\,-\,f_{ac}^{\,\,\,\,\,b}\theta^c.
\eeq
We define a riemannian structure on the group manifold via the Cartan-Killing metric tensor
\beq
\label{ckme}
g\,=\,\theta^x\otimes\theta^x+\theta^y\otimes\theta^y+\theta^z\otimes\theta^z\,=\,\theta^-\otimes\theta^++\theta^+\otimes\theta^-\,+\,\theta^z\otimes\theta^z
\eeq
associated to the Lie algebra structure of $\mathfrak{su}(2)$. From \eqref{hod} 
it is straightforward to compute that, for the Hodge duality map  one gets, with $\tau\,=\,\theta^x\wedge\theta^y\wedge\theta^z$, the following expressions on a basis
\beq
\label{hos3}
\star 1\,=\,\tau, \qquad\qquad\star \theta^a\,=\,-\dd\theta^a
\eeq
with $\star^2\,=\,1$ on $\Lambda(S^3)$.
With respect to the metric tensor \eqref{ckme} on ${\rm SU}(2)$, the Levi Civita connection acts  upon a left-invariant non holonomic basis as 
\begin{align}
\nabla_a(L_b)&=\,\,\frac{1}{2}\,f_{ab}^{\,\,\,\,\,\,s}L_s, \nn \\
\nabla_a(\theta^b)&=\,\frac{1}{2}\,f_{sa}^{\,\,\,\,\,\,b}\theta^s,
 \label{exD}
\end{align}
while  the Riemann curvature tensor gives (indices are raised and lowered via the CK-metrics)\footnote{Notice that the Christoffel symbols on such a non holonomic basis are given by the structure constants of the Lie algebra. The torsion tensor associated to it consistently (we are considering the Levi-Civita connection) vanishes, since one has $\nabla_aL_b-\nabla_bL_a=[L_a,L_b]$. }
\beq
\label{cck}
R(L_a,L_b)\theta^s\,=\,-\,\frac{1}{2}\,\nabla_{[L_a,L_b]}\theta^s.
\eeq

\subsection{The standard spin Dirac operator on ${\rm S}^3$}
\label{ss:stan}

The formalism described in the previous lines allows to define the K\"ahler-Atiyah algebra $(\Lambda({\rm S}^3), \wedge, \vee)$, which is globally defined with a Clifford product given by
\beq
\theta^a\vee\theta^b\,+\,\theta^b\vee\theta^a\,=\,2\delta^{ab}.  
\label{Cls3}
\eeq
It is well known that ${\rm S}^3$ has a spin structure,  the complexified $\Cl_{\C}({\rm S}^3, g)$ has one (up to equivalences) irreducible representation given by $\sigma\,:\,\theta^a\,\mapsto\,\sigma^a\,\in\,{\mathbb M}(\C^2)$.  On the corresponding  complex spinor bundle $S_{\C}({\rm S^3})\,=\,P_{{\rm Spin}(g)}({\rm S}^3)\times_{\sigma}\C^2$, with ${\rm Spin}(g)\,=\,{\rm SU}(2)$ one has, for the spin connection 1-form, that 
\beq
\label{oms3a}
\omega_{ab}\,=\,-\,\frac{1}{2}\,\varepsilon_{abc}\theta^c,  
\eeq
giving the following spin connection covariant derivative
\begin{align}
\delta^{\nabla}&=\,\dd\,-\,\frac{i}{4}\,\delta_{ab}\sigma^a\theta^b \nn \\
&=\,\theta^b\otimes\{L_b\,-\,\frac{i}{4}\,\delta_{ab}\sigma^a\}
\label{nas3ti}
\end{align}
so that the action of the spin Dirac operator is
\beq 
\label{spiS3D}
\Gamma(S)\,\sim\,\mathcal{F}({\rm S}^3)\,\otimes\,\C^2\,\ni\,\psi\quad\mapsto\quad\slashed D\psi\,=\,\sigma^bL_b\psi\,-\,\frac{3i}{4}\psi. 
\eeq
The spin connection Dirac operator is equivalent to  a covariant derivative -- acting upon spinors in $S_{\C}({\rm S}^3)$ --  whose connection 1-form is
\beq
\Lambda^1({\rm S}^3)\,\otimes\,{\rm End}(\C^2)\,\ni\,-\,\frac{i}{4}\theta^b\delta_{ab}\sigma^a\,\,=\,-\,\frac{i}{4}\,\left(\begin{array}{cc} \theta^3 & \theta^1-i\theta^2 \\ \theta^1+i\theta^2 & -\theta^3\end{array}\right)\,=\,\frac{1}{2}\,\gamma^{-1}\dd\gamma.
\eeq
Notice that such a connection gives the so called meron solution for Yang-Mills equations on ${\rm S}^3$ with gauge group ${\rm SU}(2)$ \cite{landimeron, dff76}. 

We are interested in explicitly computing the  spectrum of such a Dirac operator. Since ${\rm S}^3$ is compact and boundaryless, it is known that $\slashed D$ has a pure point spectrum made of discrete eigenvalues \cite{fried}; in matrix form, the spectral equation to solve for the relevant part of the operator defined in \eqref{spiS3D} is
\begin{align}
(\slashed D\,+\,3i/4)\psi\,=\,\lambda\,\psi&\qquad\Leftrightarrow\qquad(\sigma^aL_a)\psi\,=
\,\lambda\,\psi \nn \\
&\qquad\Leftrightarrow\qquad\left(\begin{array}{cc} L_z & \sqrt{2}L_- \\ \sqrt{2}L_+ & -L_z \end{array}\right)\,\left(\begin{array}{c} \psi_1 \\ \psi_2\end{array}\right)\,=\,\lambda\,\left(\begin{array}{c} \psi_1 \\ \psi_2\end{array}\right),
\label{eigequ}
\end{align}
where $\psi_i\,\in\,\mathcal{L}({\rm S}^3, \tau)$ -- $i\,=\,1,2$ -- are
the components of the spinor $\psi$. We consider the set of square integrable spinors $\Gamma(S)\,=\,
\mathcal{L}^2({\rm S}^3, \tau)\otimes\IC^2$, with 
the space of square integrable functions on ${\rm S}^3$  given by a suitable completion of the set o polynomials in the variables $(u,v,\bar{u},\bar{v})$ defined in \eqref{sug}.  The Peter-Weyl theorem proves that the Wigner functions $D^{j}_{mn}$ with $j\,=\,1/2, 1, 3/2, \ldots$ and $m,n\,\in\,(-j, -j+1, \ldots, j-1, j)$ give an orthonormal basis for $\mathcal{L}^2({\rm S}^3, \tau)$, fullfilling the relations\footnote{The $R_s$ are the right invariant vector fields on ${\rm S^3}$.}  
\begin{align}
&L^2\,D^j_{mn}\,=\,-j(j+1)\,D^j_{mn}, \nn \\
&L_z\,D^j_{mn}\,=\,in\,D^{j}_{mn}, \nn \\
&R_z\,D^j_{mn}\,=\,im\,D^j_{mn}, \label{Wigop}
\end{align}
where $L^2\,=\,\sum_{s=1}^3 L_s L_s$ is the Casimir operator (we refer to \cite{vmk} for the details of the geometry of ${\rm S}^3$) in the Lie algebra $\mathfrak{su}(2)$. Our aim is to exhibit a basis for $\Gamma(S)$ given by eigenspinors for $\slashed D$. Since the action of $\slashed D$ is written only in terms of order one left invariant derivations, $\slashed D$ commutes with $L^2$, so it is natural to consider the vector space decomposition 
$\mathcal{L}^2({\rm S}^3,\tau)\,=\oplus_jW_j$ with $W_j$ the vector space spanned by the elements $D^j_{mn}$. For example, the four elements $D^{1/2}_{mn}$ span $W_{1/2}$. Such elements are customarily cast in a $2\times 2$ matrix
\beq
D^{1/2}_{mn}\,=\,\left(\begin{array}{cc} u & \bar{v} \\ v & \bar{u} \end{array}\right), 
\label{geium}
\eeq
while the elements $D^j_{mn}$ are written,  for an arbitrary $j$, as 
\beq
D^{j}_{mn}\,=\,\left(\begin{array}{ccc} u^{2j} & \ldots & \bar{v}^{2j} \\ \ldots & \ldots & \ldots \\ v^{2j} & \ldots & \bar{u}^{2j} \end{array}\right).
\label{geiar}
\eeq
Elements in the the same column share the same eigenvalue under the action of $L_z$, while elements in the same row share the same eigenvalue under the action of $R_z$ (see \eqref{Wigop}): that is $m$ is related to the  row index, $n$ to the column index, the dimension of the matrix is $2j+1$. The action of $L_+$ shifts an element (up to suitable coefficients) to its left along the same row, i.e. $D^j_{mn}\,\mapsto \propto D^j_{m,n-1}$; the action of $L_-$ shifts an element (up to suitable coefficients) to its right along the same row, i.e. $D^j_{mn}\,\mapsto \propto D^j_{m,n+1}$.  
A basis for the Wigner functions is given by
\beq
D^j_{mn}\,=\,\propto R_z^{2m}(v^{j-n}\bar{u}^{j+n}).
\label{deff}
\eeq
In order to solve the equation \eqref{eigequ} we start by making an \emph{ansatz}, i.e. we restrict our attention to the following set of spinors 
\beq
\psi\,=\,\left(\begin{array}{c} \psi_1 \,=\,D^j_{mn} \\ \psi_2\,=\,\xi\,L_+\psi_1 \end{array}\right)  
\label{ans1}
\eeq
where $j$ is fixed, since it labels  irreducible subspaces of $\mathcal{L}^2({\rm S}^3, \tau)$ with respect to the action of ${\rm SU}(2)$, and $\xi\,\in\,\IC$. Upon inserting the condition \eqref{ans1}, the eigenvalue equation \eqref{eigequ} turns out to be  equivalent, with $n\neq-j$, to the following eigenvalue problem, 
\beq
\label{pos3}
\left(\begin{array}{cc} in & \{n^2-n-j(j+1)\}/\sqrt{2} \\ \sqrt{2} & i(1-n) \end{array}\right)\,\left(\begin{array}{c} 1 \\ \xi\end{array}\right)\,=\,\lambda\,\left(\begin{array}{c} 1 \\ \xi \end{array}\right),
\eeq
whose eigenvalue equation is\footnote{Notice that the relevant identities to obtain \eqref{pos3} are (from \eqref{lapm}):
\begin{align}
2L_-L_+\,=\,L^2-L^2_z+iL_z, \nn \\
2L_+L_-\,=\,L^2-L^2_z-iL_z, \label{icomsu}
\end{align}.}
\beq
\label{eieqq}
\lambda^2\,-\,i\lambda\,+\,j\,(j+1)\,=\,0,
\eeq
with solutions (for the eigenvalue $\lambda$ and its corresponding eigenspinor)
\begin{align}
\lambda_+\,=\,i(j+1), \qquad\qquad \psi_+\,=\,\left(\begin{array}{c} \psi_1\,=\,D^j_{mn} \\ \psi_2\,=\,-\,\frac{i\sqrt{2}}{j+n}\,L_+\psi_1 \end{array}\right) \label{pri1} \\
\lambda_-\,=\,-ij, \qquad\qquad \psi_-\,=\,\left(\begin{array}{c} \psi_1\,=\,D^j_{mn} \\ \psi_2\,=\,\,\frac{i\sqrt{2}}{j+1-n}\,L_+\psi_1 \end{array}\right). \label{pri2} 
\end{align}
Notice that, for $n=-j$, it is  $L_+D^j_{m,-j}=0$: the correct eigenspinor for $n=-j$ is, by a direct inspection,  only 
\beq
\label{tere}
\lambda_-=-ij, \qquad\qquad \psi_-\,=\,\left(\begin{array}{c}D^j_{m,-j} \\ 0\end{array}\right)
\eeq
The elements $D^j_{m,-j}$ give the first column of the matrix \eqref{geiar}. 
An element  with $j=-n$ originates, via the ansatz \eqref{ans1} only one eigenspinor, while an element with $j=n$ originates two eigenspinors corresponding to two different eigenvalues. 
Moreover, a direct inspection shows that the spinors 
\beq
\label{phize}
\psi\,=\,\left(\begin{array}{c} 0  \\ D^j_{mj}  \end{array}\right)
\eeq
are eigenspinors of the operator \eqref{eigequ}, with eigenavalue $\lambda_-=-ij$.
It is then possible to sum the above results up, and to write that $E_-(j)$, the eigenspace corresponding to the eigenvalue $\lambda_-$, has a basis given by 
\beq
E_-(j)\,=\,\left\{\left(\begin{array}{c} \psi_1\,=\,D^j_{m,n\neq\pm j} \\ \frac{i\sqrt{2}}{j+1-n}\,L_+\psi_1 \end{array}\right),\quad 
\left(\begin{array}{c} \psi_1\,=\,D^j_{m,-j} \\ 0 \end{array}\right), \quad \left(\begin{array}{c} 0  \\ \psi_2\,=\,D^j_{m,j}  \end{array}\right)
\right\},
\label{basem}
\eeq
so ${\rm dim}\,E_-(j)\,=\,2j(2j+1)+2(2j+1)$. Analogously, the eigenspace 
$E_+(j)$ corresponding to the eigenvalue $\lambda_+$ has the following basis of eigenspinors
\beq
E_+(j)\,=\,\left\{\left(\begin{array}{c} \psi_1\,=\,D^j_{m,n\neq- j} \\ -\frac{i\sqrt{2}}{j+n}\,L_+\psi_1 \end{array}\right)
\right\},
\label{basep}
\eeq
with ${\rm dim}\,E_+(j)\,=\,2j(2j+1)$. For a fixed $j$ we have that ${\rm dim}(E_-(j)\oplus E_+(j))=2(2j+1)^2$. This integer exactly gives the dimension of the subspace $W_j\otimes \IC^2$, so one can write $W_j\,=\,E_-(j)\oplus E_+(j)$.
This means that via the procedure outlined above we have been able to give a (square integrable) basis for $\Gamma(S)$ given by eigenspinors of $\slashed D$. Its spectrum is given, with the proper degeneracies,  by $\lambda=\lambda_+-3i/4=i(j-1/4)$, and $\lambda=\lambda_--3i/4=-i(j+3/4)$.

\subsection{A digression: comparing different methods to compute the spectrum of the Dirac operator}
\label{subsc}

There are many papers that address the problem of giving a spectral resolution for the spin manifold Dirac operator on ${\rm S}^3$ and in general on any sphere ${\rm S}^n$. The most cited references for this topic are \cite{fried} and \cite{baer}. The method to obtain the spectrum for $\slashed D$ and the corresponding eigenspinors is based on the notion of Killing spinor. An introduction to the subject is in \cite{fried}, it indeed traces back to works by Lichnerowitz. The aim of this digression is to sketch how this method works, and in which sense it is equivalent to the one developed above.

A spinor $\phi\,\in\,\Gamma(S)$ (without entering into a general discussion for the topic, we describe it for the geometry on the sphere ${\rm S}^3$) is a Killing spinor with Killing factor $\alpha$ if the conditions 
\beq
\label{kss3}
L_a\phi\,=\,\alpha\sigma_a\phi
\eeq 
hold. Notice that the existence of Killing spinors for a given spin structure is a non trivial problem. It is immediate to see that, if $\phi$ is a Killing spinor, than it is also an eigenspinor for $D\,=\,\sigma^aL_a$, i.e.
\beq
\label{eiphi}
\sigma^aL_a\phi\,=\,3\alpha\,\phi.
\eeq
For an arbitrary element $\psi\,\in\,\Gamma(S)$ and $f\,\in\,\mathcal{L}^2({\rm S}^3,\tau)$, the following two identities come as an easy computation from the Leibnitz rule:
\begin{align}
D(f\psi)\,&=\,(L_af)(\sigma^a\psi)+fD\psi, \label{quamo} \\
D^2(f\psi)\,&=\, i(L_af)(\sigma^a \psi)+2(L_af)(L_a\psi)+(L^2f)\psi+fD^2\psi.
\label{quaid}
\end{align}
If one considers a Killing spinor $\phi$ with factor $\alpha$, the above relations \eqref{quamo}, \eqref{quaid} give
\beq
\label{imrel}
\{D-(\alpha+i/2)\}^2(f\phi)\,=\,\{(\alpha+i/2)^2-3\alpha(2\alpha+i)+9\alpha^2\}(f\phi)+(L^2f)\phi
\eeq
If one selects $f\,\in\,W_k\,\Leftrightarrow\,L^2f\,=\,-k(k+1)f$, then the above relation takes the form
\beq
\label{almd2}
\{D-(\alpha+i/2)\}^2(f\phi)\,=\,\{(\alpha+i/2)^2-3\alpha(2\alpha+i)+9\alpha^2-k(k+1)\}f\phi
\eeq
This means that $f\phi$ is an eigenspinor of $\{D-(\alpha+i/2)\}^2$ and that a subset of the spectrum of $D$ is given by considering the square root of the relation above, namely
\begin{align}
\label{spDpa}
{\rm sp}(D)&\supset\,
(\alpha+i/2)\,\pm\,\{(\alpha+i/2)^2-3\alpha(2\alpha+i)+9\alpha^2-k(k+1)\}^{1/2}
\\
&=(\alpha+i/2)\,\pm\,\{(2\alpha-i/2)^2-k(k+1)\}^{1/2}
\end{align}
The conditions to get a Killing spinors are quite strict. The Killing spinors for the geometry we are considering are the following, both with Killing factor $\alpha\,=\,i/2$:
\beq
\label{ksex}
\phi\,=\,\left(\begin{array}{c} \bar{v} \\ u\end{array}\right), \qquad\qquad
\phi^{\prime}\,=\,\left(\begin{array}{c} \bar{u} \\ -v\end{array}\right)
\eeq
Why are these the only Killing spinors for our system? Let us  write the 
conditions \eqref{kss3} for $\phi\,=\,(\phi_1, \phi_2)$. Such are 
\beq
\label{cokil}
\left(\begin{array}{c} L_z\phi_1 \\ L_z\phi_2 \end{array}\right)\,=\,
\alpha \left(\begin{array}{c} \phi_1 \\ -\phi_2 \end{array}\right), \qquad
\left(\begin{array}{c} L_+\phi_1 \\ L_+\phi_2 \end{array}\right)\,=\,
\alpha \left(\begin{array}{c} \sqrt{2}\phi_2 \\ 0 \end{array}\right), \qquad
\left(\begin{array}{c} L_-\phi_1 \\ L_-\phi_2 \end{array}\right)\,=\,
\alpha \left(\begin{array}{c} 0 \\ \sqrt{2}\phi_1 \end{array}\right). \qquad
\eeq
The conditions $L_+\phi_2\,=\,0$ and $L_-\phi_1\,=\,0$ say that $\phi_2$ must be an element in the linear span of $\{D^j_{m,-j}\}$, i.e. the first columns of the matrices $D^j_{mn}$,  while $\phi_1$ must be spanned by the elements $\{D^j_{m,j}\}$ out of the   last columns of $D^j_{mn}$. The conditions $L_+\phi_1\,\propto\,\phi_2$ and $L_-\phi_2\,\propto\,\phi_1$ can be satisfied only if $\phi_1$ and $\phi_2$ are spanned by elements out of two columns $D^j_{mn}, \,L_+D^j_{mn}\,\propto\,D^j_{m, n-1}$.
This means that the only way to fullfill the equations \eqref{cokil} is to consider elements in $D^{1/2}_{mn}$, and the only solutions one gets are the spinors \eqref{ksex}. For $\alpha\,=\,i/2$, for the spectrum \eqref{spDpa} of $D$  one has
\beq
\label{spDal}
{\rm sp}(D)\,\supset\,i\,(1\,\pm\,(k\,+\,\frac{1}{2}))
\eeq
One can prove \cite{baer} that the expression \eqref{spDal} exhausts the spectrum of $D$, and the degeneracy of each eigenvalue can be explicitly calculated.

The comparison between the eigenspinors defined in  \eqref{basem}, \eqref{basep} and those defined as $\psi\,=\,f\,\phi$ with $L^2f\,=\,-k(k+1)f$ and $\phi$ a Killing spinor is not  immediate. In some cases one has  $W_k\,\phi\,\subset W_{j=k+1/2}$ with eigenvalue of $D$ given by $\lambda_+=i(k+3/2)=i(j+1)$; in some other cases, because of the radial condition $\bar u u+\bar v v\,=\,1$, one may have $\phi\,W_k\,\in\,W_{j=k-1/2}$. To elucidate this point, one may consider the following examples. In 
 \eqref{basem} we have seen that  
\beq
\label{eigenz}
\psi_{(1)}\,=\,\left(\begin{array}{c} u \\ 0\end{array}\right), \quad
\psi_{(2)}\,=\,\left(\begin{array}{c} v \\ 0\end{array}\right), \quad
\psi_{(3)}\,=\,\left(\begin{array}{c}  0 \\ \bar{u} \end{array}\right), \quad
\psi_{(4)}\,=\,\left(\begin{array}{c}  0 \\ \bar{v} \end{array}\right)
\eeq
are eigenspinors for $D$. It seems  that they cannot be obtained as the product $f\phi$ with a given $f$ and one of the Killing spinors. Well, they \emph{can} be suitably obtained, as the following lines show 
\begin{align}
&\psi_{(1)}\,=\,\left(\begin{array}{c} u \\ 0\end{array}\right)\,=\, vu\,\phi\,+\,u^2\,\phi^{\prime}, \nn \\
&\psi_{(2)}\,=\,\left(\begin{array}{c} v \\ 0\end{array}\right)\,=\, v^2\,\phi\,+\,uv\,\phi^{\prime}, \nn \\
&\psi_{(3)}\,=\,\left(\begin{array}{c} 0 \\ \bar u\end{array}\right)\,=\, \bar{u}^2\,\phi\,-\,\bar{u}\bar{v}\,\phi^{\prime}, \nn \\
&\psi_{(4)}\,=\,\left(\begin{array}{c} 0 \\ \bar v\end{array}\right)\,=\, \bar{u}\bar{v}\,\phi\,-\,\bar{v}^2\,\phi^{\prime}, \label{impor}
\end{align}
The eigenspinors \eqref{impor} belong to the same eigenspace for $D$, corresponding to the eigenvalue $\lambda_-\,=\,i(-k+1/2)$ with $k=1$. 

We conclude this digression by noticing that the method developed in the previous section -- which is indeed specific for ${\rm S}^3$ --  makes it easier to obtain the  the right degeneracy for the spectrum and a basis of eigenspinors, since we do not need to multiply functions and only afterwards to get the corresponding eigenvalue. With respect to the paper  \cite{cahi}, our method moreover provides the eigenspinors in terms of globally defined quantity, not depending on a local trivialisation.

\subsection{The Hodge - de Rham Dirac operator on ${\rm S}^3$}
\label{ss:HDS3}

We now adopt the algebraic approach outlined in section \ref{algspinor}, in order to define a Hodge - de Rham Dirac operator following the approach by K\"ahler. The first step is to get a Clifford algebra $\Cl_{\C}({\rm S}^3, g)$-idempotent  $P$ satisfying the relation  \eqref{compD}.
A direct proof shows that no idempotent $P$ whose range is 2-dimensional satisfies such a condition. We then consider the element
\beq
\label{P3s}
P\,=\,\frac{1}{2}\,(1\,-\,i\,\tau),  
\eeq
reading $P\vee P\,=\,P$ (since $\tau\vee\tau\,=\,-1$) and $\nabla_aP\,=\,0.$ The corresponding spinor space $I_P$ is four dimensional, with a basis given by (here $a\,=\,1, \ldots, 3$)
\begin{align}
&w_0\,=\,1-i\tau, \nn \\
& w_a\,=\,\delta_{ab}(\theta^b\,-\,\frac{i}{2}\,\varepsilon^b_{\,\,\,\,st}\theta^s\wedge\theta^t)\,=\,\delta_{ab}(\theta^b\,+\,i\dd\theta^b)\,=\,\delta_{ab}(\theta^b\,-\,i\star\theta^b).
\label{bases3}
\end{align}
The Clifford product 
\begin{align}
&\theta^a\vee w_b\,=\,i\varepsilon^{sa}_{\,\,\,\,\,\,b}w_s\,+\,\delta^a_bw_0 \nn \\
&\theta^a\vee w_0\,=\,w_a
\label{geclis}
\end{align}
gives the $\theta^a\vee\,\mapsto\,\gamma^{(a)}$ matrices representation  for the action of the Clifford algebra upon $I_P$. 
Along the basis $\{w_0, w_a\}$, one has
\beq
\gamma^{(1)}\,=\,\left(\begin{array}{cccc} 0 & 1 & 0 & 0 \\ 1 & 0 & 0 & 0 \\ 0 & 0 & 0 & -i \\ 0 & 0 & i & 0 \end{array}\right), \quad
\gamma^{(2)}\,=\,\left(\begin{array}{cccc} 0 & 0 & 1 & 0 \\ 0 & 0 & 0 & i \\ 1 & 0 & 0 & 0 \\ 0 & -i & 0 & 0 \end{array}\right), \quad
\gamma^{(3)}\,=\,\left(\begin{array}{cccc} 0 & 0 & 0 & 1 \\ 0 & 0 & -i & 0 \\ 0 & i & 0 & 0 \\ 1 & 0 & 0 & 0 \end{array}\right).
\label{gammas3}
\eeq
The action of the Levi Civita covariant derivative upon $I_P$ reads a vector potential (see  \eqref{dedie})
\begin{align}
&\nabla_jw_0\,=\,0\qquad\Rightarrow\qquad A_{j0}^{\,s}\,=\,0, \quad A_{j0}^{\,0}\,=\,0 \nn \\
&\nabla_jw_a\,=\,\frac{1}{2}\,\varepsilon^b_{\,\,\,\,ja}w_b\qquad\,\Rightarrow\qquad A_{ja}^{\,s}\,=\,\frac{1}{2}\,\varepsilon^s_{\,\,\,\,ja},\quad A_{ja}^{\,0}\,=\,0 
\label{covws3}
\end{align}
so that the corresponding connection 1-form $A$ on $I_P$ reads
\beq
\label{cdider}
\Lambda^1({\rm S}^3)\,\otimes\,{\mathbb M}(\C^4)\,\ni\,A\,=\,\frac{1}{2}\,\left(\begin{array}{cccc} 
 0 & 0 & 0 & 0 \\ 0 & 0 & -\theta^3 & \theta^2 \\ 0 & \theta^3 & 0 & -\theta^1 \\ 0 & -\theta^2 & \theta^1 & 0 
\end{array}
\right).
\eeq
The action of the 
 Hodge - de Rham Dirac operator $\D\,=\theta^a\vee\nabla_a:\,I_P\,\to\,I_P$ is
\begin{align} 
\D\,&:\,\psi^0w_0\quad\mapsto\quad \delta^{ab}(L_a\psi^0)w_b, \nn \\
&:\,\psi^aw_a\quad\mapsto\quad (L_k\psi^k)w_0\,+\,i(\varepsilon^{sk}_{\,\,\,\,\,a}L_k\psi^a\,-\,\psi^s)w_s.
 \label{Dis2hd}
\end{align}
The corresponding Laplacian operator acts as
\begin{align} 
\D^2\,&:\,\psi^0w_0\quad\mapsto\quad (L^2\psi^0)w_0, \nn \\
&:\,\psi^aw_a\quad\mapsto\quad (L^2\psi^j\,+\,\varepsilon^{jb}_{\,\,\,\,\,s}L_b\psi^s\,-\,\psi^j) w_j,
 \label{laphd3}
\end{align}
where we have denoted $L^2\,=\,\delta^{ab}L_aL_b$.

In order to study the spectrum of such a Dirac operator, one writes its action in a matrix form, so that the eigenvalue equation is
\beq
\label{Dih}
\D\psi\,=\,\left(\begin{array}{cccc} 0 & L_+ & L_z & L_- \\ L_- & L_z-i & -L_- & 0 \\ L_z  & -L_+ & -i & L_- \\ L_+ & 0 & L_+ & -i-L_z \end{array}\right)\,\left(\begin{array}{c} \psi_0 \\ \psi_- \\ \psi_z \\ \psi_+ \end{array}\right)
\,=\,\lambda\,\left(\begin{array}{c} \psi_0 \\ \psi_- \\ \psi_z \\ \psi_+ \end{array}\right)
\eeq
along the basis $w_0, w_z, w_{\pm}$. One is able to exhibit a basis of eigenspinors for $\D$ for the space of spinors identified as  $I_P\,=\,\mathcal{L}({\rm S}^3, \tau)\,\otimes \,\IC^4$ by developing ans\"atze similar to that introduced before. The spectrum of $\D$ turns out to be given by the following sectors:
\begin{enumerate}
\item for $j\neq\pm n$, one has 
\beq
\label{autov}
\lambda_{\pm}=\pm i\sqrt{j(j+1)}, \qquad\qquad 
\psi_{\pm}\,=\,\left(\begin{array}{c} \lambda_{\pm}\,D^{j}_{mn} \\ L_-D^{j}_{mn} \\ L_zD^j_{mn} \\ L_+D^j_{mn} \end{array}\right),
\eeq
\beq
\label{eigtre}
\lambda=ij, \qquad\qquad 
\psi\,=\,\left(\begin{array}{c} 0 \\ \frac{i}{j-n}\,L_{-}D^j_{mn} \\ D^{j}_{mn} \\ -\frac{i}{j+n}\,L_+D^j_{mn} \end{array}\right), 
\eeq
\beq
\label{eigv4}
\lambda=-i(j+1), \qquad\qquad 
\psi\,=\,\left(\begin{array}{c} 0 \\ -\frac{i}{j+n+1}\,L_{-}D^j_{mn} \\ D^{j}_{mn} \\ \frac{i}{j+1-n}\,L_+D^j_{mn} \end{array}\right); 
\eeq
\item for $j=n$ one has
\beq
\label{al1}
\lambda_{\pm}=\pm i\sqrt{j(j+1)}, \qquad\qquad\psi_{\pm}=\left(\begin{array}{c}\lambda_{\pm}D^j_{mj} \\ 0 \\ L_zD^j_{mj} \\ L_+D^j_{mj}\end{array}\right),
\eeq
\beq
\label{al2}
\lambda=-i(j+1), \qquad\qquad\psi=\left(\begin{array}{c} 0 \\ 0 \\ -iD^j_{mj} \\ L_+D^j_{mj}\end{array}\right),
\eeq
\beq
\label{al3}
\lambda=-i(j+1), \qquad\qquad\psi=\left(\begin{array}{c} 0 \\ 0 \\ 0 \\ D^j_{mj}\end{array}\right);
\eeq
\item for $j=-n$ one has
\beq
\label{al4}
\lambda_{\pm}=\pm i\sqrt{j(j+1)}, \qquad\qquad\psi_{\pm}\,=\,\left(\begin{array}{c}\lambda_{\pm}D^j_{m,-j} \\ L_-D^j_{m, -j} \\ L_zD^j_{m,-j} \\ 0 \end{array}\right),
\eeq
\beq
\label{al5}
\lambda=-i(j+1), \qquad\qquad\psi\,=\,\left(\begin{array}{c} 0 \\ L_-D^j_{m,-j} \\ -i\,D^j_{m, -j} \\ 0 \end{array}\right),
\eeq
\beq
\label{al6}
\lambda=-i(j+1), \qquad\qquad\psi\,=\,\left(\begin{array}{c} 0 \\ D^j_{m,-j} \\ 0 \\ 0\end{array}\right).
\eeq
\end{enumerate}
By noticing that $m$ is a degeneracy label for the eigenspinors corresponding to any given eigenvalue, we easily see  that we have a basis for  $W_j\otimes \C^4$, and then of all  $I_P$ made by eigenspinors.

\subsection{The spin manifold Dirac operator upon algebraic spinors}
\label{ss:dial}

Following the general analysis described in section \ref{ssec:spi}, we turn now to the construction of the spin manifold Dirac operator $\slashed D$ acting upon $I_P$. This means that we construct a complex spinor bundle over ${\rm S}^3$ via $W\,=\,\C^4$ since ${\rm dim}\,I_P\,=\,4$ and $\rho\,:\,\Cl_{\C}({\rm S}^3, g)\,\to\,{\rm Aut}(\C^4)$ given by $\theta^a\,\mapsto\,\gamma^{(a)}$ defined by  \eqref{geclis}. It is immediate to compute that (see \eqref{spiS3D})
\beq
\label{slDs3}
\slashed D\,=\,\gamma^a(L_a\,-\,\frac{i}{4}\,\delta_{ab}\gamma^b)\,=\,(\gamma^aL_a\,-\,\frac{3i}{4})
\eeq
and 
\begin{align}
\slashed D\,&:\,\psi^0w_0\quad\mapsto\quad-\frac{3i}{4}\,\psi^0w_0\,+\,\delta^{ks}(L_k\psi^0)w_s , \nn \\
&:\,\psi^aw_a\quad\mapsto\quad(L_k\psi^k)w_0\,+\,i(\varepsilon^{bk}_{\,\,\,\,\,a}L_k\psi^a\,-\,\frac{3}{4}\,\psi^b)w_b.
\label{slaD3s}
\end{align}
Such a Dirac operator can be seen as a covariant derivative acting upon $I_P$ with a suitable vector potential. 
From 
\begin{align}
&\frac{1}{4}\,\theta^a\eta_{ks}\Gamma_{ab}^{\,\,\,s}(\gamma^k\gamma^bw_i)\,=\,-\frac{1}{4}\,\theta^a(\varepsilon^{j}_{\,\,\,ia}w_j\,-\,\delta_{ai}w_0) \nn \\
&\frac{1}{4}\,\theta^a\eta_{ks}\Gamma_{ab}^{\,\,\,s}(\gamma^k\gamma^bw_0)\,=\,-\,\frac{i}{4}\,\theta^aw_a
\end{align}
we see that the corresponding connection 1-form 
(see \eqref{cospige}) is given by
\beq
\label{cospinma}
\Lambda^1(U_a)\,\otimes\,{\rm End}(\C^K)\,\ni\,
A\,=\,-\,\frac{1}{4}\,\left(
\begin{array}{cccc} 0 & i\theta^1 & i\theta^2 & i\theta^3 \\ i\theta^1 & 0 & \theta^3 & -\theta^2 \\ i\theta^2 & -\theta^3 & 0 & \theta^1 \\ i\theta^3 & \theta^2 & -\theta^1 & 0   
\end{array}
\right).
\eeq
The Hodge - de Rham Dirac operator and the spin Dirac operator can both be seen as covariant derivatives acting upon the same spinor space $I_P$, with different vector potentials, since the one in \eqref{cospinma} differs from the one in \eqref{cdider}.
The corresponding Laplacian is
\begin{align}
\slashed D^2\,&:\,\psi^0w_0\quad\mapsto\quad(L^2\psi^0\,-\,\frac{9}{16}\psi^0)w_0\,-\,\frac{i}{2}\,(L_k\psi^0)\delta^{ks}w_s
 \nn \\
&:\,\psi^aw_a\quad\mapsto\quad -\,\frac{i}{2}\,(L_k\psi^k)w_0\,+\,(L^2\psi^j\,+\,\frac{1}{2}\, 
\varepsilon^{js}_{\,\,\,\,\,a}L_s\psi^a\,-\,\frac{9}{16}\,\psi^j)w_j.
\label{slalap3s}
\end{align}
As we did for the Dirac operator we have previously analysed, we turn now to the problem of determining the spectrum of the operator $\slashed D\,+3i/4\,=\,\gamma^aL_a$ acting upon the algebraic spinors $I_P$ as in \eqref{slaD3s}. The matrix form of the corresponding eigenvalue equation is 
\beq
\label{Diha}
\gamma^aL_a\psi\,=\,\left(\begin{array}{cccc} 0 & L_+ & L_z & L_- \\ L_- & L_z & -L_- & 0 \\ L_z  & -L_+ & 0 & L_- \\ L_+ & 0 & L_+ & -L_z \end{array}\right)\,\left(\begin{array}{c} \psi_0 \\ \psi_- \\ \psi_z \\ \psi_+ \end{array}\right)
\,=\,\lambda\,\left(\begin{array}{c} \psi_0 \\ \psi_- \\ \psi_z \\ \psi_+ \end{array}\right)
\eeq
along the basis $w_0, w_z, w_{\pm}$. Notice that the difference between \eqref{Dih} and \eqref{Diha} resides in the constant terms along the diagonal.  Exploring  ans\"atze close to the one already considered,  one can check that the spectrum for the operator \eqref{Diha} can be written as follows: 
\begin{enumerate}[(1)]
\item for $j\neq\pm n$, 
\beq
\begin{array}{c} \lambda_+=i(j+1) \\ \lambda_-=-ij\end{array} \qquad\qquad \psi_{\pm}=\left(\begin{array}{c} D^j_{mn} \\ \frac{1}{\lambda_{\pm}-i}\,L_-D^j_{mn} \\ -\frac{n}{1+i\lambda_{\pm}}\,D^j_{mn} \\ 
\frac{1}{\lambda_{\pm}-i}\,L_+D^j_{mn} \end{array}\right), \qquad\qquad \tilde{\psi}_{\pm}=\left(\begin{array}{c} -\frac{n}{1+i\lambda_{\pm}}\,D^j_{mn} \\ \frac{1}{-\lambda_{\pm}+i}\,L_-D^j_{mn} \\ D^j_{mn} \\ 
\frac{1}{\lambda_{\pm}-i}\,L_+D^j_{mn} \end{array}\right) ;
\label{se1b}
\eeq
\item for $n=-j$, since $L_+D^j_{m,-j}=0$,  one has
\beq
\lambda_-=-ij, \qquad\qquad\psi_{(1)}=\left(\begin{array}{c} D^{j}_{m,-j} \\ 0 \\ D^{j}_{m,-j} \\ 0\end{array}\right), \qquad 
\psi_{(2)}=\left(\begin{array}{c} D^{j}_{m,-j} \\ i\,L_-D^{j}_{m,-j} \\ 0 \\ 0\end{array}\right), \qquad \psi_{(3)}=\left(\begin{array}{c} 0 \\ D^{j}_{m,-j} \\ 0  \\ 0\end{array}\right),
\label{se2b}
\eeq
\beq
\lambda_+=i(j+1), 
\qquad\qquad\qquad \psi_{(4)}=\left(\begin{array}{c} D^{j}_{m,-j} \\ -ij^{-1}\,D^{j}_{m,-j} \\ -D^j_{m,-j}  \\ 0\end{array}\right);
\label{se3b}
\eeq
\item for $n=j$, with $L_-D^j_{mj}=0$, one has
\beq
\lambda_-=-ij, \qquad
\qquad\qquad\tilde{\psi}_{(1)}=\left(\begin{array}{c} D^{j}_{mj} \\ 0 \\ -D^{j}_{mj}  \\ 0 \end{array}\right), \qquad 
\tilde{\psi}_{(2)}=\left(\begin{array}{c} D^{j}_{mj} \\ 0 \\ 0 \\  -i\,L_+D^{j}_{mj} \end{array}\right), \qquad \tilde{\psi}_{(3)}=\left(\begin{array}{c} 0 \\ 0 \\ 0 \\ D^{j}_{mj} \end{array}\right),
\label{se4b}
\eeq
\beq
\lambda_+=i(1+j), \qquad\qquad\qquad\tilde{\psi}_{(4)}=\left(\begin{array}{c} D^j_{mj} \\ 0 \\ D^j_{mj} \\ -ij^{-1}D^j_{mj} \end{array}\right). 
\label{se5b}
\eeq
\end{enumerate}
Upon counting the multiplicities of the above eigenvalues, one can conclude that they give the whole spectrum of the operator \eqref{Diha}.

\subsection{The K\"ahler-Dirac operator on ${\rm S}^2$}
\label{ssp2}
It is well known that the quotient of the right action of a ${\rm U}(1)$ subgroup upon ${\rm S}^3$ is the basis ${\rm S}^2$ of the principal Hopf bundle. If we choose $L_z$ as the infinitesimal generator of such a principal action, then we identify $L_z$ as the vertical vector field of the fibration and   realise the exterior algebra $\Lambda({\rm S}^2)$ over the basis as the set of horizontal and ${\rm U}(1)$-equivariant forms over ${\rm S}^3$, i.e.
\beq
\Lambda({\rm S}^2)\,=\,\{\phi\,\in\,\Lambda({\rm S}^3)\,:\,L_z\phi\,=\,0, \,i_{L_z}\phi\,=\,0\},
\label{as2}
\eeq
so we may write
\beq
\label{ef2}
\Lambda({\rm S}^2)\,\ni\,f\,+\,c_-\theta^-\,+\,c_+\theta^+\,+\,ih\,\theta^{-}\wedge\theta^+
\eeq
under the conditions that $(f,c_-, c_+,  h)$ are  elements  in $\mathcal{F}(\rm S^3)$  
fulfilling the conditions
\begin{align}
&L_zf\,=\,L_zh\,=\,0 \nn \\
&L_zc_{\pm}\,=\,{\mp}ic_{\pm}.
\label{coes2}
\end{align}
Notice that the conditions on the second line  originate from  \eqref{livf} and \eqref{l1pm}, reading  $L_z\theta^{\pm}\,=\,i\theta^{\pm}$. We are  assuming  $\Lambda(S^2)$ to have complex coefficients.

Since the exterior algebra over ${\rm S}^2$ is realized as a subset of the exterior algebra over ${\rm S}^3$, we wonder how the K\"ahler-Dirac operator defined on ${\rm S}^3$ acts upon differential forms over ${\rm S}^2$. From \eqref{lpm}-\eqref{ckme} it is straightforward to compute that, on ${\rm S}^3$, one has 
\begin{align}
&\dd\theta^-\,=\,-i\,\theta^-\wedge\theta^z,  \nn \\
&\dd\theta^+\,=\,i\,\theta^+\wedge\theta^z,  \nn \\
&\dd\theta^z\,=\,i\,\theta^-\wedge\theta^+
\label{dds3}
\end{align}
while, for the Hodge duality map on $\rm S^3$ one gets, with $\tau\,=\,i\,\theta^-\wedge\theta^+\wedge\theta^z$, the following expressions on a basis
\beq
\label{hos3}
\star 1\,=\,\tau, \qquad\qquad\star \theta^a\,=\,\dd\theta^a
\eeq
with $\star^2\,=\,1$ on $\Lambda(\rm S^3)$. One can explicitly calculate that, if $f\,\in\,\mathcal{F}({\rm S}^3), \,L_zf\,=\,0$, then
\beq
\label{errdir}
\mathcal{D}(f\theta^-\wedge\theta^+)\,\notin \,\Lambda({\rm S}^2).
\eeq
The action of the Hodge - de Rham Dirac operator on ${\rm S}^3$ does not induce a meaningful operator when its domain is restricted to $\Lambda({\rm S}^3)$. The formalism described so far allows indeed to introduce a meaningful Hodge - de Rham Dirac operator on the euclidean sphere ${\rm S}^2$.

The set $\Lambda({\rm S}^2)$ is a free bimodule over $\mathcal{F}({\rm S}^2)$ of dimension 8. 
This claim has an easy proof. Following   \cite{lapro}
we consider the array of elements in $\mathcal{F}(S^3)$ (see \eqref{sug})
\beq
\varphi\,=\,(u^2, \,\sqrt{2}uv, \,v^2), \qquad\qquad \bar{\varphi}\,=\,(\bar{u}^2, \,\sqrt{2}\bar{u}\bar{v}, \,\bar{v}^2)
\label{mopro}
\eeq
with (componentwise)  $\sum_j\varphi_j\bar{\varphi}_j\,=\,1$. From \eqref{livf} it is immediate to see that 
\begin{align}
&L_z\,c_-\,=\,i\,c_-\quad\Leftrightarrow\quad c_-\,=\,\sum_j\alpha_j\bar{\varphi}_j, \nn \\
&L_z\,c_+\,=\,-i\,c_+\quad\Leftrightarrow\quad c_+\,=\,\sum_j\beta_j\varphi_j, 
\label{modufo}
\end{align}
with $L_z\alpha_j\,=\,0;\,\,L_z\beta_j\,=\,0$.

We now explore how to obtain irreducible actions of the Dirac operator upon a set of algebraic spinors on ${\rm S}^2$.
The subset $\Lambda({\rm S}^2)$ is a Clifford subalgebra of $(\Lambda({\rm S}^3), \wedge, \vee)$, as one can compute
\begin{align}  
&c_{\pm}\theta^{\pm}\,\vee\, c^{\prime}_{\pm}\theta^{\pm}\,=\,0, \nn \\
&c_{-}\theta^{-}\,\vee\,c_+\theta^{+}\,=\,c_-\theta^-\,\wedge\, c_+\theta^+\,+\,1 \nn \\
&c_-\theta^-\,\vee\,c_+\theta^+\,+\,c_+\theta^+\,\vee\,c_-\theta^-\,=\,2.
\end{align}
These relations allows to prove the following identity:
\begin{align}
\label{cl2s}
&(f+c_-\theta^-+c_+\theta^++ih\theta^-\wedge\theta^+)\vee
(f^{\prime}+c^{\prime}_-\theta^-+c^{\prime}_+\theta^++ih^{\prime}\theta^-\wedge\theta^+)\,=\,\\
&\qquad(ff^{\prime}+c_-c^{\prime}_++c_-^{\prime}c_+-hh^{\prime})\,+\,(fh^{\prime}+f^{\prime}h-ic_-c_+^{\prime}+ic_+c_-^{\prime})i\theta^-\wedge\theta^+ \,+ \nn \\
&\qquad\qquad(fc_-^{\prime}+f^{\prime}c_-+ihc_-^{\prime}-ih^{\prime}c_-)\theta^-\,+\,(fc_+^{\prime}+f^{\prime}c_+-ihc_+^{\prime}+ih^{\prime}c_+)\theta^+. \nn 
\end{align}
We look now for idempotent elements in $\Lambda({\rm S}^2, \wedge, \vee)$, that is we look for elements satisfying the identity $P\vee P=P$. From the relation above, one easily sees that $\Lambda({\rm S}^2)\,\ni\,P\,=\,f=c_-\theta^-+c_+\theta^++ih\theta^-\wedge\theta^+$ is a projector if and only if 
\beq
\label{copro2}
f\,=\,\frac{1}{2}, \qquad 2c_-c_+\,=\,\frac{1}{4}\,+\,h^2.
\eeq
We stress that solutions for such equations exist only if we complexify the Clifford algebra on ${\rm S^3}$ and on ${\rm S}^2$. 
We consider the element 
\beq
\label{prs2}
P\,=\,\frac{1}{2}\,(1\,-\,\theta^-\wedge\theta^+)\,=\,1\,-\,\frac{1}{2}\,\theta^-\vee\theta^+
\eeq
which satisfies the identity $P\vee P=P$. From \eqref{cl2s} it is straightforward to prove that
\begin{align}
&c_-\theta^-\,\vee\,P\,=\,c_-\theta^-, \nn \\
&c_+\theta^+\,\vee\,P\,=\,0, \nn \\
&\theta^-\vee\theta^+\vee P\,=\,0
\label{alide}
\end{align}
so that 
\beq
\label{ips}
I_P\,=\,\{\Lambda(S^2)\,\ni\,\psi\,=\,f\,+\,c_-\theta^-\,-\,\frac{1}{2}\,f\,\theta^-\vee\theta^+\}
\eeq
where $(f, \,c_-)$ are complex valued elements in $\mathcal{F}({\rm S}^3)$ satisfying the conditions \eqref{coes2}. As a module on $\mathcal{F}({\rm S}^2)$, $I_P$ is free and 4-dimensional. In order to introduce a Dirac operator we need to introduce a suitable Hodge duality on $\Lambda({\rm S}^2)$. The way to do that is to restrict the scalar product defined on $\Lambda({\rm S}^3)$ as in \eqref{scaproe}  to its subalgebra $\Lambda({\rm S}^2)$, and to prove that with respect to such a restriction a meaningful Hodge $\star_{{\rm S}^2}$ can be defined\footnote{See the footnote \ref{notap} and the references \cite{gial, ale-hodge}, where a Hodge duality  for the non commutative sphere ${\rm S}_q^2$
has been defined via a frame bundle approach which can be directly adapted
to the classical case we are considering.}, reading
\begin{align}
\star_{{\rm S}^2}(f)&=\,if\theta^-\wedge\theta^+,  \nn \\
\star_{{\rm S}^2}(c_-\theta^-)&=\,-ic_-\theta^-,  \nn \\
\star_{{\rm S}^2}(c_+\theta^+)&=\,ic_+\theta^+,  \nn \\
\star_{{\rm S}^2}(if\theta^-\wedge\theta^+)&=\,f 
\label{stars2}
\end{align}
We can write \eqref{ips} as
\beq
\label{spi2s}
I_P\,=\,\{\Lambda({\rm S}^2)\,\ni\,\psi\,=\,f\,\psi_1\,+\,(\sum_j\alpha_j\bar{\varphi}_j)\theta^-\}
\eeq
with $\psi_1\,=\,\frac{1}{2}\,(1-\theta^-\wedge\theta^+)$. We have
\begin{align}
\D(f\psi_1)\,=\,(L_-f)\theta^-\,=\,\sum_j\{\varphi_j\,L_-f\}(\bar{\varphi}_j\theta^-), \nn \\
\D(\sum_j\alpha_j\bar{\varphi}_j\theta^-)\,=\,\{L_+(\sum_j\alpha_j\bar{\varphi}_j)\}(1-\theta^-\wedge\theta^+)\,=\,2\,(L_+\sum_j\alpha_j\bar{\varphi}_j)\psi_1,
\label{dboh}
\end{align}
with 
\begin{align}
L_z(\sum_j\{\varphi_j\,L_-f\}\,=\,0, \nn \\
L_z(L_+\sum_j\alpha_j\bar{\varphi}_j)\,=\,0.
\label{idsf}
\end{align}
Along the basis $(\psi_1, \bar{\varphi}_j\theta^-)$ of $I_P$, the action of the Dirac operator has the following matrix form:
\beq
\label{dima2ss}
\D\,=\,\left(\begin{array}{cccc} 0  & 2(\bar{\varphi}_1L_+\,+\,(L_+\bar{\varphi}_1))  & 2(\bar{\varphi}_2L_+\,+\,(L_+\bar{\varphi}_2)) 
& 2(\bar{\varphi}_3L_+\,+\,(L_+\bar{\varphi}_3)) \\ 
\varphi_1L_- & 0 & 0 & 0 \\
\varphi_2L_- & 0 & 0 & 0 \\
\varphi_3L_- & 0 & 0 & 0 
\end{array}\right)
\eeq
where the terms $\bar{\varphi}_jL_+\,$ are meaningful order 1 differential operators upon functions on ${\rm S}^2$, while the terms  $(L_+\bar{\varphi}_j)$ are to be read as meaningful  multiplication operators upon functions on ${\rm S}^2$ (see \eqref{idsf}). 

We are interested in computing the spectrum of the operator  $\D$ defined above. We define a spinor $\psi\,=\,(\psi_0, \,\psi_1, \,\psi_2, \,\psi_3)$, so that the eigenvalue equation $\D\psi\,=\,\lambda\psi$ turns out to be equivalent to the following componentwise equations, with $k=1,2,3$,
\begin{align}
&\sum_k\,2\{\psi_k(L_+\bar{\varphi}_k)\,+\,\bar{\varphi}_k(L_+\psi_k)\}\,=\,\lambda\psi_0, \label{eqe1} \\
&\varphi_k(L_-\psi_0)\,=\,\lambda\,\psi_k
\label{eqe2}
\end{align}
Since one already knows (Lichnerowitz identity) that  $\lambda\,\neq\,0$, one casts the expression $\psi_k\,=\,
\lambda^{-1}\varphi_k(L_-\psi_0)$ into \eqref{eqe1} and, after some algebra, obtains
\beq
2\,L_+L_-\psi_0\,=\,\lambda^2\psi_0
\label{eqe3}
\eeq
From the Lie algebra relations for the rotation group one has the identity $2L_+L_-\,=\,L^2\,-\,L^2_z\,-\,iL_z$. From \eqref{coes2} we characterise the set $\mathcal{L}^2({\rm S}^2, i_z\tau)$  of square integrable functions on ${\rm S}^2$ as the ${\rm Ker}\,L_z$ acting upon square integrable functions on ${\rm S}^3$. This means that each $\psi_0\,=\,D^j_{m,0}$ is a solution of \eqref{eqe3} with $\lambda^2\,=\,-j(j+1)$. The spectrum resolution of  the Dirac operator $\D$ turns out to be
\beq
\lambda_{\pm}\,=\,\pm\,i\,\sqrt{j(j+1)},\qquad\qquad \psi_{\pm}\,=\,\left(\begin{array}{c} \psi_0\,=\,D^j_{m,0} \\ \psi_1\,=\,\lambda^{-1}_{\pm}\varphi_1\,L_-\psi_0 \\  
\psi_2\,=\,\lambda^{-1}_{\pm}\varphi_2\,L_-\psi_0 \\
\psi_3\,=\,\lambda^{-1}_{\pm}\varphi_3\,L_-\psi_0 
\end{array}\right)
\label{espis2}
\eeq
Notice that the above expression is meaningful only for integer $j\,=\,1,2,3,\ldots$.




\subsection{A local description of the K\"ahler-Dirac operator on ${\rm S}^2$}
\label{sss:local}

The aim of this section is to report some explicit calculations developed within  the local manifold formalism for ${\rm S}^2$ equipped with the usual Riemannian metric tensor. We consider a local chart for ${\rm S}^2$ with a coordinate system given   by  $\theta\,\in\,(0, \pi),\quad \varphi\,\in\,[0, 2\pi)$ and the metric tensor given by 
\beq
\label{riemg}
g\,=\,\dd\theta\otimes\dd\theta+\sin^2\theta\,\dd\varphi\otimes\dd\varphi\,=\,\ct\otimes\ct\,+\,\hat{\varphi}
\otimes\hat{\varphi},
\eeq
where the \emph{zweibein} for such a metric structure is given by $\hat\theta\,=\,\dd\theta, \quad\hat\varphi\,=\,\sin\theta\,\dd\varphi$, so we have a local K\"ahler-Atiyah algebra $(\Lambda(S^2), \wedge, \vee)$, 
with  the volume form $\tau\,=\,\ct\wedge\hat{\varphi}\,=\,\sin\theta\,\dd\theta\wedge\dd\varphi$. 

The set of projectors $P$ with $P\vee P=P$ satisfying the integrability condition \eqref{compD} is given by
\beq
\label{probeta}
P_{\pm}(\beta)\,=\,\frac{1}{2}\,(1\pm i\tau)\,+\,\beta(\ct\pm i\cp)
\eeq
with $\beta\,\in\,\C$. The set of algebraic spinors, that is the set $$\Lambda({\rm S}^2)\,\supseteq\,I_{\pm}(\beta)\,=\,\{\psi\,:\,\psi\,\vee\,P_{\pm}(\beta)\,=\,\psi
\}$$ is a 2-dimensional $\mathcal{F}({\rm S}^2)$-bimodule with basis given by
\beq
\label{baspin}
\psi^{(\pm)}_1\,=\,1\pm i\tau, \qquad\qquad\psi^{(\pm)}_2\,=\,\ct\pm i\hat{\varphi}.
\eeq
Notice that $I_{\pm}(\beta)$ does not depend on $\beta$. The $\Pin({\rm S}^2, g)$ group is given by elements 
\beq 
\epsilon\,=\,A\ct\,+\,B\hat{\varphi}, \qquad\qquad A,\,B\,\in\,\C\,:\,A^2+B^2\,=\,\pm1
\label{ping}
\eeq
and the notion of equivalence in the set of projectors given by \eqref{eqP} reads
\begin{align}
\epsilon\vee\epsilon\,=\,1&\qquad\Rightarrow\qquad\epsilon\vee P_{\pm}(\beta)\vee\epsilon^{-1}\,=\,P_{\mp}((A\mp iB)^2\beta), \nn \\
\epsilon\vee\epsilon\,=\,-1&\qquad\Rightarrow\qquad\epsilon\vee P_{\pm}(\beta)\vee\epsilon^{-1}\,=\,P_{\mp}((\pm iA+ B)^2\beta), \label{equPP}
\end{align}
while the corresponding transformation $\psi\,\mapsto\,\psi\vee\epsilon$ for  the set of spinors  reads $I_{\pm}\vee\epsilon\,=\,I_{\mp}$. Upon spinors in $I_-$ along the basis $\{\psi^{(-)}_{1,2}\}$ the action of the K\"ahler-Dirac operator has the matrix form
\beq
\label{Dis2}
\mathcal{D}\,=\,\left(\begin{array}{cc} 0 & \frac{\cos\theta}{\sin\theta}\,+\,\del_{\theta}\,-\,\frac{i}{\sin\theta}\,\del_{\varphi} \\ 
\del_{\theta}\,+\,\frac{i}{\sin\theta}\,\del_{\varphi} & 0 \end{array}\right).
\eeq
To prove this result one easily checks that the Clifford action of the generators $\ct, \,\hat\varphi$ upon the above basis in $I_-$ reads the 2-dimensional irreducible representation given by
\beq 
\ct\,\vee\,\mapsto\, \sigma_x,\qquad\qquad\qquad \hat\varphi\,\vee\,\mapsto\,\sigma_y,
\label{reps2}
\eeq
(these identities give the first order differential terms of the operator $\mathcal{D}$)
while the identities
\beq
\mathcal{D}\,:\,\psi_1^{(-)}\,\mapsto\,0, \qquad\qquad \mathcal{D}\,:\,\psi_2^{(-)}\,\mapsto\,\frac{\cos\theta}{\sin\theta}\,\psi_1^{(-)}
\label{Dino}
\eeq
are responsible for the curvature term in \eqref{Dis2}. 

Which is the action of the spin Dirac operator on sections of the usual spinor bundle over ${\rm S}^2$? If one identifies the spinor space by $I_-$, following \cite{fof}, one gets 
\beq
\label{spinDs2}
\slashed D\,=\,\left(\begin{array}{cc} 0 & \frac{\cos\theta}{2\sin\theta}\,+\,\del_{\theta}\,-\,\frac{i}{\sin\theta}\,\del_{\varphi} \\ 
\frac{\cos\theta}{2\sin\theta}\,+\,\del_{\theta}\,+\,\frac{i}{\sin\theta}\,\del_{\varphi} & 0 \end{array}\right).
\eeq
This operator coincides with the one written in \cite{abri}. One easily sees that $\mathcal{D}$ and $\slashed D$ are not gauge related. Moreover, a local description of both operators is given in \cite{varilly}, where $\mathcal{D}$ is not presented as acting upon algebraic spinors.

\subsection*{Acknowledgements}  
We started to study this topic under the suggestion and the guidance of our advisor, Beppe Marmo, to whom we are gratefully indebted for the precious insights he gave us. We should like to thank Patrizia Vitale, Juan Manuel P\'erez - Pardo and Erlend Grong for the many interesting discussions together. 
This paper originated during the Workshop on \emph{Quantum Physics: foundations and applications} (CHEP -- Bangalore 
2016). We should like to thank the organizers and all the participants for the stimulating atmosphere we experienced. 
We gratefully acknowledge the support of the INFN, of Norbert Poncin and the University of Luxembourg. This paper has been evolved and then completed  while we were visiting the ICMAT (Madrid) and the Brookhaven National Laboratory (Upton, NY), which we thank.


\begin{thebibliography}{99}

\bibitem{abri} A.A. Abrikosov Jr., {\it Dirac operator on the Riemann sphere}, arXiv:hep-th/0212134.

\bibitem{baer} C. B\"ar, {\it The Dirac operator on space forms of positive curvature}, J. Math. Soc. Japan 48 (1996) 69-83.


\bibitem{balaimmi} A.P. Balachandran, G. Immirzi, J. Lee, P. Presnajder, {\it Dirac operators on coset spaces}, J. Math. Phys. 44 (2003) 4713-4735. 

\bibitem{cahi} R. Camporesi, A. Higuchi, {\it On the eigenfunctions of the Dirac operator on spheres and real hyperbolic spaces}, J. Geom. and Phys. 20 (1996) 1-18.

\bibitem{dff76} V. de Alfaro, S. Fubini, G. Furlan, {\it A new classical solution of the Yang-Mills field equations}, Phys. Lett. B65 (1976) 163-166. 

\bibitem{fof} J. Figueroa-O'Farrill, {\it Spin Geometry}, unpublished, lecture Notes avalaible at http://empg.maths.ed.ac.uk/Activities/Spin

\bibitem{VFG} J.M. Gracia-Bond\'ia, J.C. Varilly, H. Figueroa, {\it Elements of Noncommutative Geometry}, Birkh\"auser (2000).  

\bibitem{fried} T. Friedrich, {\it Dirac operators in Riemannian geometry}, AMS 2000.

\bibitem{graf} W. Graf, {\it Differential forms as spinors}, Annales I.H.P, section A, 29 (1978)  85-192.


\bibitem{ka} E. K\"ahler, {\it Der innere Differentialkalk\"ul}, Rendiconti di Matematica vol. 21 (1962) 425-523.


\bibitem{lapro} G. Landi, {\it Projective modules of finite type and monopoles over $S^2$}, J. Geom and Phys. 37 (2001) 47-62.

\bibitem{landimeron} G. Landi, {\it Spinor and gauge connections over oriented spheres}, in Proceedings of the 7th Italian Conference on General Relativity and Gravitational Physics, World Scientific (1987),  287-291.  

\bibitem{gial} G. Landi, A. Zampini, {\it Calculi, Hodge operators and Laplacians on a quantum Hopf fibration}, Rev. Math. Phys. 23 (2011) 575-613.

\bibitem{landsman} N.P. Landsman, {\it Notes on Noncommutative Geometry}, unpublished lecture notes avalaible at www.math.ru.nl/~landsman/ncg2010.pdf. 

\bibitem{lami} H.B. Lawson, M.-L. Michelson, {\it Spin geometry}, Princeton University Press 1989.

\bibitem{nicolae} L.I. Nicolaescu, {\it Lectures on the geometry of manifolds}, World Scientific 1996.

\bibitem{plymen} R.J. Plymen, {\it Strong Morita equivalence, spinors and symplectic spinors}, J. Oper. Theory 16 (1986) 305-324.

\bibitem{schroeder} H. Schr\"oder, {\it On the definition of geometric Dirac operators}, math.dg/0005239.

\bibitem{tucker} R.W. Tucker, {\it A Clifford calculus for physical field theories}, in {\it Clifford algebras and their applications in mathematical physics}, NATO ASI Series vol.183 (1985).

\bibitem{varilly} J. Varilly, {\it Dirac operators and spectral geometry}, 2006: 
unpublished lecture notes 
avalaible at 
https://www.impan.pl/swiat-matematyki/notatki-z-wyklado$\sim$/varilly$\_$dosg.pdf


\bibitem{vmk} D.A. Varshalovic, A.N. Moskalev, V.K. Khersonskii, {\it Quantum theory of angular momentum}, World Scientific 1988.


\bibitem{ale-hodge} A. Zampini, {\it Hodge duality operators on left covariant exterior algebras over two and three dimensional quantum spheres}, Rev. Math. Phys. 25 (2013) 9-38.

\end{thebibliography}
\end{document}